\documentclass[12pt,a4paper]{article}

\usepackage{epsf}
\usepackage{latexsym,amssymb,euscript}
\usepackage[dvips]{graphicx}
\usepackage[numbers,sort&compress]{natbib}
\usepackage{amsmath}
\usepackage{slashed}
\usepackage{booktabs}
\usepackage{hyperref}
\usepackage{braket}
\usepackage{bbold}
\usepackage{graphics}
\usepackage{graphicx}
\usepackage{caption}
\usepackage{subcaption}
\usepackage[titletoc]{appendix}
\graphicspath{{./figures/}}
\hypersetup{
 linktocpage = true,
 urlcolor = purple,
 colorlinks = true,
 linkcolor = purple,
 anchorcolor = purple,
 citecolor = purple,
 pdfstartview = {XYZ null null 1.25} 
           }
\usepackage[left=3cm, right=2cm]{geometry}
\usepackage{pstricks}
\usepackage{color}
\usepackage{float}

\newcommand{\tarr}{
\begin{array}}
\newcommand{\earr}{\end{array}}
\newcommand*\rfrac[2]{{}^{#1}\!/_{#2}} 

\newcommand{\drv}{{\rm d}}
\newcommand{\LL}{$\left( \Lambda\text{-}\Lambda \right)$}
\newcommand{\LJ}{$\left( \Lambda\text{-jet} \right)$}
\newcommand{\CnNLA}{{\cal C}_n^{\rm NLA}}
\newcommand{\CnDGLAP}{{\cal C}_n^{\rm DGLAP}}

\begin{document}

\begin{titlepage}

\begin{center}
{\LARGE \bf Diffractive production of $\boldmath{\Lambda}$ hyperons \\ in the high-energy limit of strong interactions}
\end{center}

\vskip 0.5cm

\centerline{
Francesco~Giovanni~Celiberto$^{1,2*}$,
Dmitry~Yu.~Ivanov$^{3,4\S}$, 
and Alessandro~Papa$^{5,6\ddagger}$}

\vskip .6cm

\centerline{${}^1$ {\sl Dipartimento di Fisica, Universit\`a degli Studi di Pavia, I-27100 Pavia, Italy}}
\vskip .2cm
\centerline{${}^2$ {\sl INFN, Sezione di Pavia, I-27100 Pavia, Italy}}
\vskip .2cm
\centerline{${}^3$ {\sl Sobolev Institute of Mathematics, 630090 Novosibirsk,
   Russia}}
\vskip .2cm
\centerline{${}^4$ {\sl Novosibirsk State University, 630090 Novosibirsk,
   Russia}}
\vskip .2cm
\centerline{${}^6$ {\sl Dipartimento di Fisica, Universit\`a della Calabria}}
\centerline{\sl I-87036 Arcavacata di Rende, Cosenza, Italy}
\vskip .2cm
\centerline{${}^5$ {\sl Istituto Nazionale di Fisica Nucleare, Gruppo collegato
      di Cosenza}}
\centerline{\sl I-87036 Arcavacata di Rende, Cosenza, Italy}
\vskip 2cm

\begin{abstract}
 We propose the study of the inclusive production of two $\Lambda$ hyperons or a single $\Lambda$-particle in association with a jet, featuring high transverse momenta and large separation in rapidity, as a probe channel of the resummation of energy logarithms in the QCD perturbative series. We give predictions, shaped on kinematic ranges typical of CMS and of the backward CASTOR detector, for cross section and azimuthal-correlation moments between the two emitted objects, showing how considering the tag of $\Lambda$ baryons eases the comparison between theoretical results and experimental data in the phenomenological context of semi-hard reactions.
\end{abstract}

\vskip .5cm

$^{*}${\it e-mail}:
\href{mailto:francescogiovanni.celiberto@unipv.it}{francescogiovanni.celiberto@unipv.it}

$^{\S}${\it e-mail}:
\href{mailto:d-ivanov@math.nsc.ru}{d-ivanov@math.nsc.ru}

$^{\ddagger}${\it e-mail}:
\href{mailto:alessandro.papa@fis.unical.it}{alessandro.papa@fis.unical.it}

\end{titlepage}


\section{Introduction}
\label{introduction}

The study of basic properties and decay channels of baryon particles is widely recognized as a landmark to deepen our knowledge of strong interactions. Inside the baryon family, a very interesting class of particles is represented by \emph{hyperons}, namely hadrons whose lowest Fock state contains one or more strange valence quarks. 
The apparent anomaly observed in decay times of kaons and hyperons inspired physicists of the middle of the last century first to introduce the \emph{strangeness} conservation law~\cite{Pais:1952zz,Nakano:1953zz,Nishijima:1955gxk,Gell-Mann:1956iqa,GellMann:1955jx}, then to organize hadrons via the \emph{eightfold way}~\cite{GellMann:1961ky,Neeman:1961jhl}. 
Nowadays, studies on hyperons are relevant in the search for \emph{CP}-violation signatures and exotic states, as well as in spectroscopy and spin analyses.

$\Lambda$ hyperons have been subject of intense investigation in the last years (see Refs.~\cite{Soffer:2003qm,Boros:1999da} for an overview). 
As an example, \emph{single-spin asymmetries} in unpolarized hadron collisions have been first observed in the case of $\Lambda$ emissions, then confirmed in the detection of several other hyperons. Here, although QCD dynamics at partonic level forbids any sizeable asymmetry~\cite{Dharmaratna:1989jr,Dharmaratna:1996xd}, $\Lambda$ particles produced in the beam fragmentation region are largely polarized along the direction orthogonal to the production plane.
This effect still represents an unsolved puzzle~\cite{Soffer:1999ww} and many studies on the connection between spin and non-perturbative distributions have been carried out so far. It is even more puzzling that such effect is not
confirmed in electron-proton and electron-nucleus collisions: data for
semi-inclusive deep-inelastic scattering (SIDIS) from NOMAD~\cite{Astier:2000ax}
and ZEUS~\cite{Chekanov:2006wz} show spin asymmetries compatible with zero, though within large uncertainties.  Analyses on collinear \emph{polarized} $\Lambda$ fragmentation functions (FFs) and on the \emph{transverse-momentum-dependent} (TMD)\footnote{For a review on TMD factorization see, \emph{e.g.}, Refs.~\cite{Rogers:2015sqa,Diehl:2015uka,Echevarria:2015uaa} and references therein.}  \emph{polarizing} $\Lambda$ FF have been made in Ref.~\cite{Kotzinian:1997vd,deFlorian:1998ba,Anselmino:2000ga,Anselmino:2001ps} and in Refs.~\cite{Mulders:1995dh,Anselmino:2000vs,Anselmino:2001js,Anselmino:2019cqd}, respectively. A first extraction of the polarizing $\Lambda$ FF can be found in Ref.~\cite{DAlesio:2020wjq}.
In Ref.~\cite{Ellis:1995fc} it was highlighted that measurements of $\Lambda$ polarization states in deep-inelastic configurations may shed light on dynamical mechanisms invoked to explain the \emph{proton spin puzzle}~\cite{Ashman:1987hv}.

In the studies listed above $\Lambda$ particles are detected in final-state configurations featuring relatively small rapidities (or rapidity intervals) and large Feynman variables (say, $x_F~\gtrsim~0.2$).

When outgoing particles are emitted in forward rapidity regions of detectors, low-$x$ dynamics naturally comes into play.
In the \emph{Regge limit}, namely where the center-of-mass energy squared, $s$, is definitely larger than the Mandelstam~$t$, large energy logarithms enter the perturbative series in the strong coupling, $\alpha_s$, with a power increasing with the order. This spoils the convergence of pure fixed-order analyses in collinear factorization, calling for an all-order resummation action. 
The Balitsky--Fadin--Kuraev--Lipatov (BFKL) formalism~\cite{BFKL_1,BFKL_2,BFKL_3,BFKL_4} provides us with a rigorous way to resum to all orders these large-energy logarithms both in the leading approximation (LLA), which means $(\alpha_s \ln s)^n$ type logarithms,
and in the next-to-leading approximation (NLA), which means $\alpha_s (\alpha_s \ln s)^n$ type ones.
In the BFKL approach, the imaginary part the amplitude of a hadronic process reads as an elegant convolution between two impact factors, depicting the transition from each parent particle to the corresponding final-state object, and a gluon Green's function.
The latter is process independent and it is controlled by an integral evolution equation, its kernel being known in the next-to-leading-order approximation (NLO)~\cite{Fadin:1998py,Ciafaloni:1998gs,Fadin:1998jv,Fadin:2000kx,Fadin:2004zq,Fadin:2005zj}. Impact factors depend both on the initial and on the final-state particle~\cite{Fadin:1999de,Fadin:1999df,Ciafaloni:1998kx,Ciafaloni:1998hu,Ciafaloni:2000sq}, therefore they need to be calculated process by process and only few of them are know with NLO accuracy.

Diffractive \emph{semi-hard} reactions~\cite{Gribov:1984tu}, namely diffractive processes which exhibit the following scale hierarchy, $\sqrt{s} \gg \{ Q \} \gg \Lambda_{\rm QCD}$ ($\{ Q \}$ is a (set of) process-characteristic hard scale(s), while $\Lambda_{\rm QCD}$ the is QCD scale), are widely recognized as golden channels to test the high-energy limit of strong interactions via the BFKL resummation. This stems from the fact that large final-state rapidities or rapidity distances, typical of diffractive final states, increase the weight of contributions proportional to $\ln (s)$ in the semi-hard regime.
Two distinct classes of diffractive semi-hard final states can be singled out: single forward emissions and forward/backward two-particle detections. To the first class, the exclusive leptoproduction of a light vector meson~\cite{Anikin:2009bf,Anikin:2011sa,Besse:2012ia,Besse:2013muy,Bolognino:2018rhb,Bolognino:2018mlw,Bolognino:2019bko,Celiberto:2019slj,Bolognino:2019pba}, the inclusive hadroproduction of a bottom-quark~\cite{Chachamis:2015ona}, the exclusive photoproduction of a  quarkonium~\cite{Bautista:2016xnp,Garcia:2019tne} and the inclusive forward Drell--Yan dilepton production~\cite{Motyka:2014lya,Brzeminski:2016lwh,Celiberto:2018muu} belong.
On the other hand, a wide range of processes pertaining to the second class has been proposed in the last years (see Refs.~\cite{Celiberto:2017ius,Celiberto:2020wpk} for a review): the two-meson exclusive leptoproduction~\cite{Ivanov:2004pp,Ivanov:2005gn,Ivanov:2006gt,Enberg:2005eq},
the total cross section of two deeply-virtual photons~\cite{Ivanov:2014hpa},
the inclusive hadroproduction of two jets emitted with high transverse momenta
and large rapidity separation (Mueller--Navelet configuration~\cite{Mueller:1986ey}),
for which several phenomenological analyses have been realized so
far~\cite{Marquet:2007xx,Colferai:2010wu,Caporale:2012ih,Ducloue:2013wmi,Ducloue:2013bva,Caporale:2013uva,Ducloue:2014koa,Caporale:2014gpa,Colferai:2015zfa,Caporale:2015uva,Ducloue:2015jba,Celiberto:2015yba,Celiberto:2015mpa,Celiberto:2016ygs,Celiberto:2016vva,Caporale:2018qnm,Chachamis:2015crx}, the inclusive multi-jet
hadroproduction~\cite{Caporale:2015vya,Caporale:2015int,Caporale:2016soq,Caporale:2016vxt,Caporale:2016xku,Celiberto:2016vhn,Caporale:2016djm,Caporale:2016lnh,Caporale:2016zkc}, the inclusive emission of two light-charged hadrons~\cite{Ivanov:2012iv,Celiberto:2016hae,Celiberto:2016zgb,Celiberto:2017ptm}, $J/\Psi$-jet~\cite{Boussarie:2017oae},
hadron-jet~\cite{Bolognino:2018oth,Bolognino:2019yqj,Bolognino:2019cac}, Higgs-jet~\cite{Celiberto:2020tmb},
Drell--Yan-jet~\cite{Golec-Biernat:2018kem,Deak:2018obv} and heavy-flavored di-jet
photo-~\cite{Celiberto:2017nyx,Bolognino:2019ouc} and
hadroproduction~\cite{Bolognino:2019yls}.

The BFKL resummation still represents a powerful tool to improve our understanding of the proton structure at small-$x$. First, it allowed to define and study an \emph{unintegrated gluon distribution} (UGD)~\cite{Forshaw:1997dc}, written as a convolution of the gluon Green's function and the non-perturbative proton impact factor (for a study on DIS structure functions, see Ref.~\cite{Hentschinski:2012kr}).
Then, it gave the chance to improve the description of collinear parton distribution functions (PDFs) at
NLO and next-to-NLO (NNLO) through the inclusion of NLA resummation effects~\cite{Ball:2017otu,Abdolmaleki:2018jln,Bonvini:2019wxf}. Ultimately, it permitted to predict the small-$x$ behavior of TMD gluon distributions~\cite{Bacchetta:2020vty}.

\begin{figure}[t]
 \centering
 \includegraphics[scale=0.45]{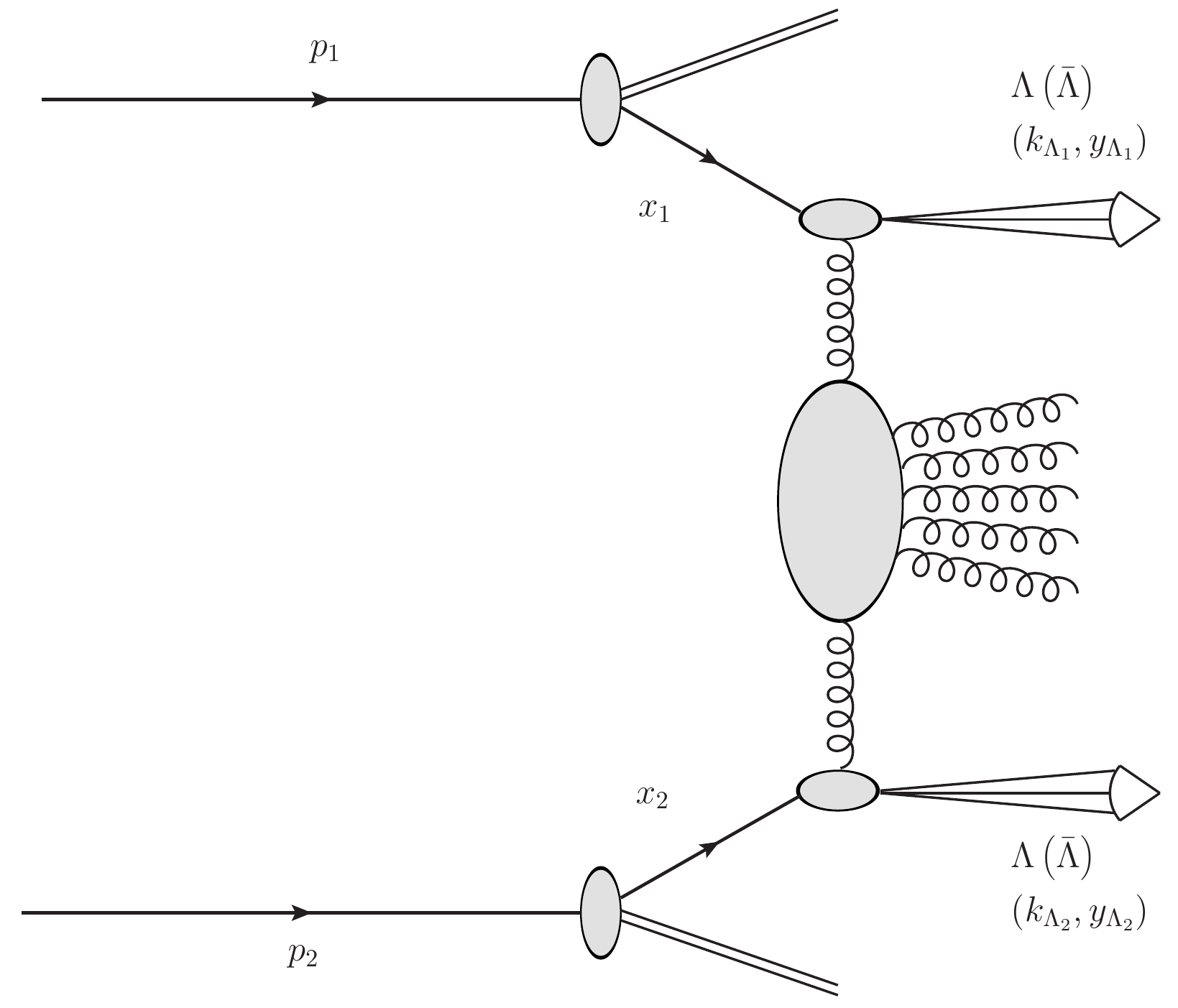}
 \hspace{0.75cm}
 \includegraphics[scale=0.45]{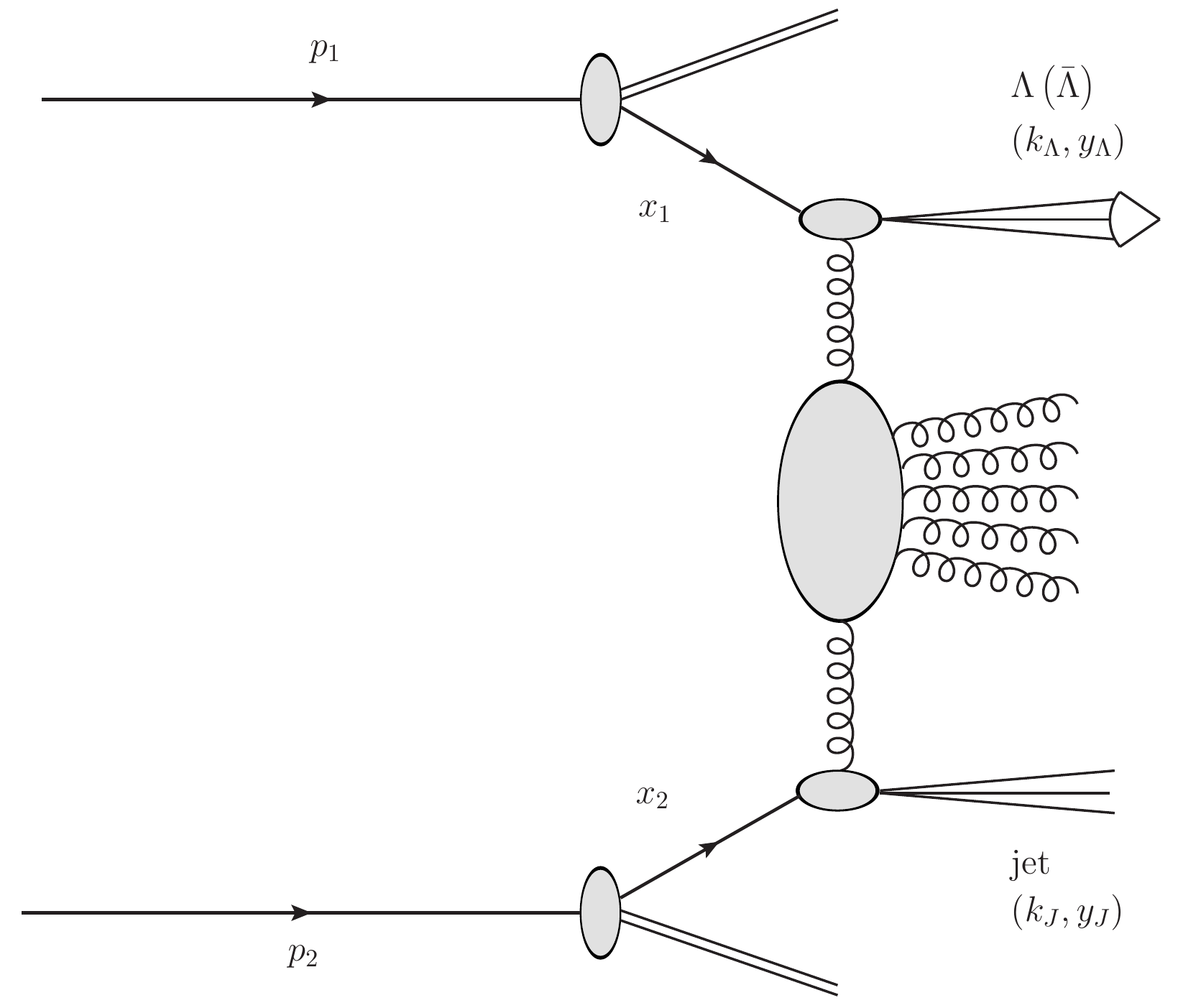}
 \\ \vspace{0.25cm}
 a) {\LL} channel \hspace{4.50cm}
 b) {\LJ} channel
 \caption[]
 {Diagrammatic representation of the inclusive diffractive {\LL} and of the {\LJ} hadroproduction (panels a) and b), respectively). Final-state objects are produced in the fragmentation region of the corresponding parent hadrons, together with secondary gluon emission in the central-rapidity range.}
 \label{fig:LL-LJ}
\end{figure}

In this paper we propose the inclusive emission of $\Lambda$ hyperons in diffractive semi-hard configurations as an additional channel to probe the high-energy resummation in the kinematic ranges typical of current and upcoming experimental studies at the LHC. In particular, we focus on final states featuring the emission of a forward (backward) $\Lambda$ particle accompanied by another $\Lambda$ (panel a) of Fig.~\ref{fig:LL-LJ}) or by a jet (panel b) of Fig.~\ref{fig:LL-LJ}) tagged in backward (forward) directions. The final-state inclusiveness is warranted by the emission of undetected hard gluons strongly ordered in rapidity. Similar configurations have been extensively investigated with NLA accuracy in the context of di-hadron~\cite{Celiberto:2017ptm,Celiberto:2020wpk} and hadron-jet~\cite{Bolognino:2018oth,Celiberto:2020wpk} correlations. Nevertheless, the detection of $\Lambda$ baryons brings several benefits.
First, it allows us to quench minimum-bias effects better than emissions of lighter charged hadrons~\cite{Bolognino:2018oth}, thus easing the comparison with experimental data.
Moreover, it affords us the opportunity to access naturally asymmetric final-state kinematic configurations\footnote{The rise of collinear contaminations due to relatively small rapidity intervals in the {\LL} channel can be compensated by considering {\LJ} configurations, with the jet tagged by CMS or by the CASTOR ultra-backward detector.}, an essential ingredient to discriminate BFKL from other resummations~\cite{Celiberto:2015yba,Celiberto:2015mpa,Celiberto:2020wpk}. Then, it provides us with a complementary channel to further probe and constrain collinear FFs describing the production mechanism of unpolarized $\Lambda$ hyperons, which currently represents an important challenge in the enhancement of our knowledge of QCD. Generally, it enriches the collection of semi-hard reactions which can serve as a testfield of the dynamics of strong interactions in the high-energy limit.

Our task is to estimate the feasibility of such studies, calculating values of cross sections and azimuthal angle correlations. In what follows we will use the MOM scheme for the strong coupling renormalization with Brodsky--Lepage--Mackenzie (BLM) optimization~\cite{Brodsky:1996sgBrodsky:1997sdBrodsky:1998knBrodsky:2002ka} of the renormalization scale fixing, because earlier such approach to BFKL resummation proved to be successful in the description of recent LHC data on Mueller--Navelet jet production. 

\section{Inclusive diffractive production of $\boldmath{\Lambda}$ hyperons}
\label{theory}

A general formula for the final states under investigation can be presented as:
\begin{equation}
\label{process}
 {\rm proton}(p_1) + {\rm proton}(p_2) \to \Lambda(k_1, y_1) + X + P_2(k_2 , y_2) \;,
\end{equation}
where a $\Lambda$ hyperon\footnote{In our numerical calculations below we will always present results for the sum of baryon and antibaryon production cross sections.} is always detected in association with another $\Lambda$ or a jet, $P_2 \equiv \{\Lambda, {\rm jet}\}$ (panels a) and b) of Fig.~\ref{fig:LL-LJ}). Both final-state particles feature large transverse momenta, $|\vec k_{1,2}| \equiv \kappa_{1,2} \gg \Lambda_{\rm QCD}$, and consistent distance in rapidity, $ Y \equiv y_1 - y_2$. Furthermore a secondary, inclusive hadronic subsystem, $X$, is mostly produced in more central regions of rapidity.
The protons' momenta, $p_{1,2}$ are taken as Sudakov vectors satisfying $p^2_{1,2} = 0$ and $(p_1 p_2) = s/2$,  allowing for a suitable decomposition of the momenta of the produced objects:
\begin{equation}
\label{sudakov}
 k_{1,2} = x_{1,2} p_{1,2} + \frac{\vec k_{1,2}^2}{x_{1,2} s} p_{2,1} + k_{{1,2 \perp}} \;, \qquad k_{1,2 \perp}^2 = - \vec k_{1,2}^2 \equiv - \kappa_{1,2}^2 \; .
\end{equation}

In the center-of-mass system, the final-state longitudinal momentum fractions, $x_{1,2}$, are related to the respective rapidities through the expressions
$y_{1,2} = \pm \frac{1}{2}\ln\frac{x_{1,2}^2 s}
{\kappa_{1,2}^2}$,
so that $\drv y_{1,2} = \pm \frac{\drv x_{1,2}}{x_{1,2}}$, 
and $\Delta Y = y_1 - y_2 = \ln\frac{x_1 x_2 s}{\kappa_1\kappa_2}$, here the
space part of the four-vector $p_{1 \parallel}$ being taken positive.

Using collinear factorization, an overall expression for the cross sections of our interest reads
\begin{equation}
\label{sigma_collinear}
 \frac{\drv \sigma}{\drv x_1 \drv x_2 \drv^2 \vec k_1 \drv^2 \vec k_2}
 =\sum_{s,t=q,{\bar q},g}\int_0^1 \drv x_1 \int_0^1 \drv x_2\ f_s\left(x_1,\mu_{F1}\right)
 \ f_t\left(x_2,\mu_{F2}\right)
 \frac{\drv {\hat\sigma}_{s,t}\left(\hat s,\mu_{F1,2}\right)}
 {\drv x_1 \drv x_2 \drv^2 \vec k_1 \drv^2 \vec k_2}\;,
\end{equation}
where the ($s,t$) indices run over the parton kinds 
(quarks $q = u, d, s, c, b$;
antiquarks $\bar q = \bar u, \bar d, \bar s, \bar c, \bar b$; 
or gluon $g$), $f_{s,t}\left(x, \mu_{F1,2} \right)$ are the initial proton PDFs; 
$x_{1,2}$ stand for the longitudinal fractions of the partons involved in the hard
subprocess, whereas $\mu_{F1}$~($\mu_{F2}$) is the factorization scale characteristic of the fragmentation region of the upper (lower) parent proton in panels of Fig.~\ref{fig:LL-LJ};
$\drv \hat\sigma_{s,t}\left(\hat s,\mu_{F1,2} \right)$ denotes
the partonic cross section and $\hat s \equiv x_1x_2s$ is the squared center-of-mass energy of the partonic collision.

\subsection{High-energy resummed cross section} 
\label{cross_section}

In the BFKL approach the cross section is suitably given 
(see Ref.~\cite{Caporale:2012ih} for the details of the derivation) as the Fourier sum of the azimuthal coefficients, $\CnNLA$, having so
\begin{equation}
 \label{dsigma_Fourier}
 \frac{\drv \sigma}{\drv y_1 \drv y_1 \drv \kappa_1 \drv \kappa_2 \drv \vartheta_1 \drv \vartheta_2} =
 \frac{1}{(2\pi)^2} \left[{\cal C}^{\rm NLA}_0 + 2 \sum_{n=1}^\infty \cos (n \varphi)\,
 \CnNLA \right]\, ,
\end{equation}
where $\vartheta_{1,2}$ are the azimuthal angle of the tagged particles and $\varphi = \vartheta_1 - \vartheta_2 - \pi$.
A NLA-BFKL consistent formula for the azimuthal-angle averaged cross section, ${\cal C}^{\rm NLA}_0$, and for the other coefficients, ${\cal C}^{\rm NLA}_{n \ge 1}$, can be presented in the momentum renormalization (MOM) scheme (whose definition is related to the three-gluon vertex, an essential ingredient of the BFKL resummation) as
\[
\CnNLA = \frac{x_1 x_2}{\kappa_1 \kappa_2}\, 
\int_{-\infty}^{+\infty} \drv \nu \,
 \left( \frac{\hat s}{\kappa_1 \kappa_2} \right)^{\frac{N_c}{\pi} \alpha^{\rm MOM}_s(\mu_R)\left[\chi(n,\nu)
+\frac{N_c}{\pi} \alpha^{\rm MOM}_s(\mu_R)\left(\bar \chi(n,\nu) +\frac{T^{\rm conf}}
{3}\chi(n,\nu)\right)\right]}
\]
\[
\left[ \alpha^{\rm MOM}_s (\mu_R) \right]^2
c_1(n,\nu,\kappa_1, x_1)[c_2(n,\nu,\kappa_2,x_2)]^*
\]
\begin{equation}
\label{Cn}
\times \,
 \left\{1 + \alpha^{\rm MOM}_s(\mu_R)\left[\frac{\bar c_1(n,\nu,\kappa_1,x_1)}{c_1(n,\nu,\kappa_1, x_1)}
 +\left[\frac{\bar c_2(n,\nu,\kappa_2, x_2)}{c_2(n,\nu,\kappa_2,x_2)}\right]^*
 +\frac{2T^{\rm conf}}{3} \right] \right\} \, .
\end{equation}
Here, the MOM expression for the running coupling can be obtained from the corresponding one in the $\overline{\rm MS}$ scheme via
\begin{equation}
 \alpha^{\rm MOM}_s (\mu_R) = 
 - \frac{\pi}{2 (T^{\beta} + T^{\rm conf})}
 \left( 1 - \sqrt{1 + 4 \, \alpha^{(\overline{\rm MS})}_s (\mu_R) \frac{T^{\beta} + T^{\rm conf}}{\pi}} \right)
 \label{as_MOM} \;,
\end{equation}
with
\[
\label{T_bc}
T^{\beta}=-\frac{\beta_0}{2}\left( 1+\frac{2}{3}I \right) \; ,
\]
\begin{equation}
T^{\rm conf}= \frac{C_A}{8}\left[ \frac{17}{2}I +\frac{3}{2}\left(I-1\right)\xi
+\left( 1-\frac{1}{3}I\right)\xi^2-\frac{1}{6}\xi^3 \right] \; ,
\end{equation}
where $I = -2 \int_0^1 \drv u \frac{\ln \left( u \right)}{u^2 - u + 1} \simeq 2.3439$ and $\xi$ is a gauge parameter, fixed at zero in the following. 
Furthermore, $N_c$ is the color number and $C_A = N_c$, $\beta_0 = \rfrac{11}{3} N_c - \rfrac{2}{3}n_f$
is the first coefficient of the QCD $\beta$-function, with $n_f$ is the active-flavor number,
\begin{equation}
 \chi \left( n, \nu \right) = 
 2 \psi \left (1 \right) - 
 2 \, {\rm Re} \, \{ \psi \left (\rfrac{n}{2} + \rfrac{1}{2} + i\nu \right) \}
\end{equation}
stands for the leading-order (LO) BFKL characteristic function, $\bar \chi(n,\nu)$ is the eigenvalue of NLA BFKL
kernel~\cite{Kotikov:2000pm} and its expression can be found, {\it e.g.} in Eq.~(23) of Ref.~\cite{Caporale:2012ih}, $c_{1,2}(n,\nu)$ depict the LO forward particle impact factor in the ``so-called'' $(n,\nu)$-repre\-sen\-ta\-tion. The $\Lambda$-emission is described at LO by the light-hadron impact factor, which reads
\[
c_\Lambda(n,\nu,\kappa_\Lambda,x_\Lambda) 
= 2 \sqrt{\frac{C_F}{C_A}}
(\kappa_\Lambda^2)^{i\nu-1/2}\,\int_{x_\Lambda}^1\frac{\drv \alpha}{\alpha}
\left( \frac{\alpha}{x_\Lambda} \right)
^{2 i\nu-1} 
\]
\begin{equation}
\label{LOHIF}
 \times \left[\frac{C_A}{C_F}f_g(\alpha)D_g^\Lambda\left(\frac{x_\Lambda}{\alpha}\right)
 +\sum_{s=q,\bar q}f_s(\alpha)D_s^\Lambda\left(\frac{x_\Lambda}{\alpha}\right)\right] \;,
\end{equation}
where $D_i^\Lambda(x_\Lambda / \alpha)$ is the FF for the $\Lambda$-particle ``generated'' in the final state with longitudinal fraction $x_\Lambda$, from hadronization of the parton $i$ with longitudinal fraction $\alpha$.
Analogously, the tagged jet is depicted by the corresponding LO impact factor
\begin{equation}
 \label{LOJIF}
 c_J(n,\nu,\kappa_J,x_J) =  2 \sqrt{\frac{C_F}{C_A}}
 (\kappa_J^2)^{i\nu-1/2}\,\left(\frac{C_A}{C_F}f_g(x_J)
 +\sum_{t=q,\bar q}f_t(x_J)\right) \;.
\end{equation}
The remaining functions are the $\beta_0$-independent parts of the NLO impact factor corrections, $\bar c_{\Lambda,J}(n,\nu,|\kappa_{\Lambda,J}|, x_{\Lambda,J})$
, their formulas being given in Eqs.~(4.58)-(4.65) of Ref.~\cite{Ivanov:2012iv}
and in Eq.~(36) of Ref.~\cite{Caporale:2012ih}, respectively. For brevity, we do not show here the dependencies of the impact factors on the renormalization and factorization scales.

The peculiar form of Eq.~(\ref{Cn}) turns up as an outcome of the BLM~\cite{Brodsky:1996sgBrodsky:1997sdBrodsky:1998knBrodsky:2002ka} scale-optimization method\footnote{In this paper we limit ourselves to give the final expression for the azimuthal coefficients in the MOM scheme with BLM optimization. The interested reader can find the formal derivation in Section~3 of Ref.~\cite{Caporale:2015uva}.}, which prescribes to take the ``optimal'' renormalization scale, $\bar \mu$, as the value that lead to the vanishing of the non-conformal, $\beta_0$-dependent terms in the expression for the observable of interest. The concurrent presence of these terms both in the NLA BFKL gluon Green's function and in the NLO impact-factor corrections makes the BLM scale non-universal, but rather dependent on energy~\cite{Caporale:2015uva}. The operational criterion for the BLM scale setting is to fix $\bar \mu$ as the solution of the following integral equation 
\[
C^{(\beta)}_n (s, Y)
=
\int \drv \Phi_{1,2}
\int\limits^{\infty}_{-\infty} \drv \nu \,
\left( \frac{\hat s}{\kappa_1 \kappa_2} \right)^{\frac{N_c}{\pi} \alpha^{\rm MOM}_s(\bar \mu) \chi(n, \nu)}
c_1(n,\nu,\kappa_1, x_1)[c_2(n,\nu,\kappa_2,x_2)]^*
\]
\begin{equation}
\times \,
 \left[{\cal F}(\nu, \bar \mu) + \ln \left( \frac{\hat s}{\kappa_1 \kappa_2} \right) \frac{N_c}{\pi} \alpha^{\rm MOM}_s(\bar \mu) \, \chi(n,\nu) \left(- \frac{\chi(n,\nu)}{4} + \frac{{\cal F}(\nu, \bar \mu)}{2} \right) \right] = 0 \, ,
 \label{Cn_beta}
\end{equation}
where
\begin{equation}
\label{fnu_hat}
{\cal F}(\nu, \bar \mu) = i \left[ \frac{1}{2} \frac{\drv}{\drv \nu} \ln\left(\frac{c_1(n,\nu,\kappa_1, x_1)}{[c_2(n,\nu,\kappa_2,x_2)]^*}\right) + \ln \left( \kappa_1 \kappa_2 \right) \right] + \frac{5}{3} + 2 \ln \frac{\bar \mu}{\kappa_1 \kappa_2} - 2 - \frac{4}{3} I
\end{equation}
and $\Phi_{1,2}$ is the phase space of the particles produced in the final state (see Section~\ref{phase_space}).
We choose $\bar \mu$ in the form of a multiple of the geometric mean of the two natural scales of the process, $\bar \mu = d_R \sqrt{\kappa_1 \kappa_2}$, and look for the values of $d_R$ which solve Eq.~(\ref{Cn_beta}). Then, we plug the found value for the renormalization scale (which can be arbitrarily chosen within NLA accuracy) into the expression of the azimuthal coefficients, and we set $\mu_{F_{1,2}} = \mu_R$, as assumed by most of the existent PDF parametrizations.

We compare our NLA BFKL results with fixed-order predictions based on an effective high-energy DGLAP calculation (for more details on its derivation, see Refs.~\cite{Celiberto:2015yba,Celiberto:2020wpk}), where the $\CnDGLAP$ azimuthal coefficients are introduced as truncation to the ${\cal O}(\alpha_s^3)$ order of the corresponding NLA BFKL ones, $\CnNLA$, up to the inclusion of terms beyond the LO accuracy. This permits to pick the leading-power asymptotic features of a pure NLO DGLAP description, discarding at the same time those terms which are dampened by inverse powers of the energy of the partonic subprocess.
Our DGLAP expression reads
\begin{equation}
\label{DGLAP_Cn}
 \CnDGLAP \equiv \frac{x_1 x_2}{\kappa_1 \kappa_2}
 \int_{-\infty}^{+\infty} \drv \nu \, \left[ \alpha^{\rm MOM}_s (\mu_R) \right]^2
 c_1(n,\nu,\kappa_1, x_1)[c_2(n,\nu,\kappa_2,x_2)]^*\,
\end{equation}
\[
 \times \, \left\{1+\alpha^{\rm MOM}_s(\mu_R) \left[\frac{C_A}{\pi} \ln \left( \frac{\hat s}{\kappa_1 \kappa_2} \right) \chi(n,\nu) 
 \right. \right.
 \]
 \[
 \left. \left.
 + \, \frac{\bar c_1(n,\nu,\kappa_1,x_1)}{c_1(n,\nu,\kappa_1, x_1)}
 + \left[\frac{\bar c_2(n,\nu,\kappa_2, x_2)}{c_2(n,\nu,\kappa_2,x_2)}\right]^* 
 + \frac{2T^{\rm conf}}{3}
 \right]
 \right\} \;,
\]
where the BFKL exponentiated kernel has been replaced by its expansion up to terms proportional to $\alpha_s(\mu_R)$.

\subsection{Final-state observables}
\label{phase_space}

We integrate the azimuthal coefficients, ${\cal C}_n$, over the phase space of the two final-state objects, by keeping the mutual rapidity distance, $\Delta Y$, fixed. One has
\begin{equation}
 \label{Cn_int}
 C_n =
 \int_{\kappa_1^{\rm inf}}^{\kappa_1^{\rm sup}} \drv \kappa_1
 \int_{\kappa_2^{\rm inf}}^{\kappa_2^{\rm sup}} \drv \kappa_2
 \int_{y_1^{\rm inf}}^{y_1^{\rm sup}} \drv y_1
 \int_{y_2^{\rm inf}}^{y_2^{\rm sup}} \drv y_2
 \, \,
 \delta (\Delta Y - (y_1 - y_2))
 \, \,
 {\cal C}_n\left(\kappa_1, \kappa_2, y_1, y_2 \right)
 \, ,
\end{equation}
where the integration over the rapidity of the second particle, $y_2$, will be removed by imposing the delta condition, $\delta(\Delta Y - (y_1 - y_2))$. 
Here, ${\cal C}_n$ and $C_n$ indistinctly refer to the corresponding NLA BFKL calculations (Eq.~(\ref{Cn})) or the high-energy DGLAP ones (Eq.~(\ref{DGLAP_Cn})). 
We consider realistic kinematic configurations, suggested by recent experimental analyses at the LHC. In our study, $\Lambda$ particles are detected (both {\LL} and {\LJ} channels) in the symmetric rapidity range from~$-2.0$~to~2.0 and feature transverse momenta larger than 10~GeV, according to the typical CMS measurements for the $\Lambda_b$ baryon~\cite{Chatrchyan:2012xg}, that we use as a proxy for the detection of $\Lambda$  hyperons. Then, two possibilities for the jet emission ({\LJ} channel) are considered: a) the symmetric \textit{CMS-jet} case~\cite{Khachatryan:2016udy}, with $|y_J| < 4.7$ and 35 GeV~$< \kappa_J < \kappa_{J,{\rm CMS}}^{\max} = 60$~GeV; b) the ultra-backward (with respect to CMS rapidity acceptances) \textit{CASTOR-jet} configuration~\cite{CMS:2016ndp}, namely when the jet is tagged by the CASTOR detector with $-6.6 < y_J < -5.2$ and 10~GeV~$< \kappa_J < \kappa_{J,{\rm CST}}^{\max} \simeq 17.68$~GeV.
The upper bounds of $\kappa_J$ are constrained by requiring that $x_J \le 1$, whereas the value adopted for the upper bound of $\kappa_\Lambda$, $\kappa_{\Lambda,{\rm CMS}}^{\rm max} = 21.5$~GeV, is constrained by the lower cutoff of the FF sets (see below).
The final-state observables under investigation are the
the $\varphi$-averaged cross section, $C_0$, and the azimuthal ratios, $R_{nm} \equiv C_{n}/C_{m}$. Among
them, the $R_{n0}$ ratios have an immediate physical interpretation, being the correlation moments, $\langle \cos n \varphi \rangle$, while the ones without indices equal to zero correspond to ratios of cosines, $\langle \cos n \varphi \rangle / \langle \cos m \varphi \rangle$~\cite{Vera:2006un,Vera:2007kn}. We study the $\Delta Y$-dependence of our observables at $\sqrt{s} = 13$~TeV and in the range, $\Delta Y \ge 1.5$.

\subsection{Numerical analysis and discussion}
\label{results}

The numerical analysis was performed using the {\tt JetHad}~\cite{Celiberto:2020wpk} modular interface, suited to the analysis of inclusive semi-hard reactions.
PDFs were calculated through {\tt MMHT2014nlo} parameterizations~\cite{Harland-Lang:2014zoa} as provided by {\tt LHAPDFv6.2.1}~\cite{Buckley:2014ana}, while {\tt AKK2008nlo} FF routines~\cite{Albino:2008fy} were selected to describe $\Lambda$-baryon emissions.
Error bands in our plots return the numerical uncertainty rising from the multidimensional integration over the final-state phase space. All calculations were done in the MOM scheme, while a two-loop running-coupling choice with $\alpha_s\left(M_Z\right)=0.11707$ and
dynamic-flavor threshold was made.

\begin{figure}[t]
\centering

 \includegraphics[scale=0.52,clip]{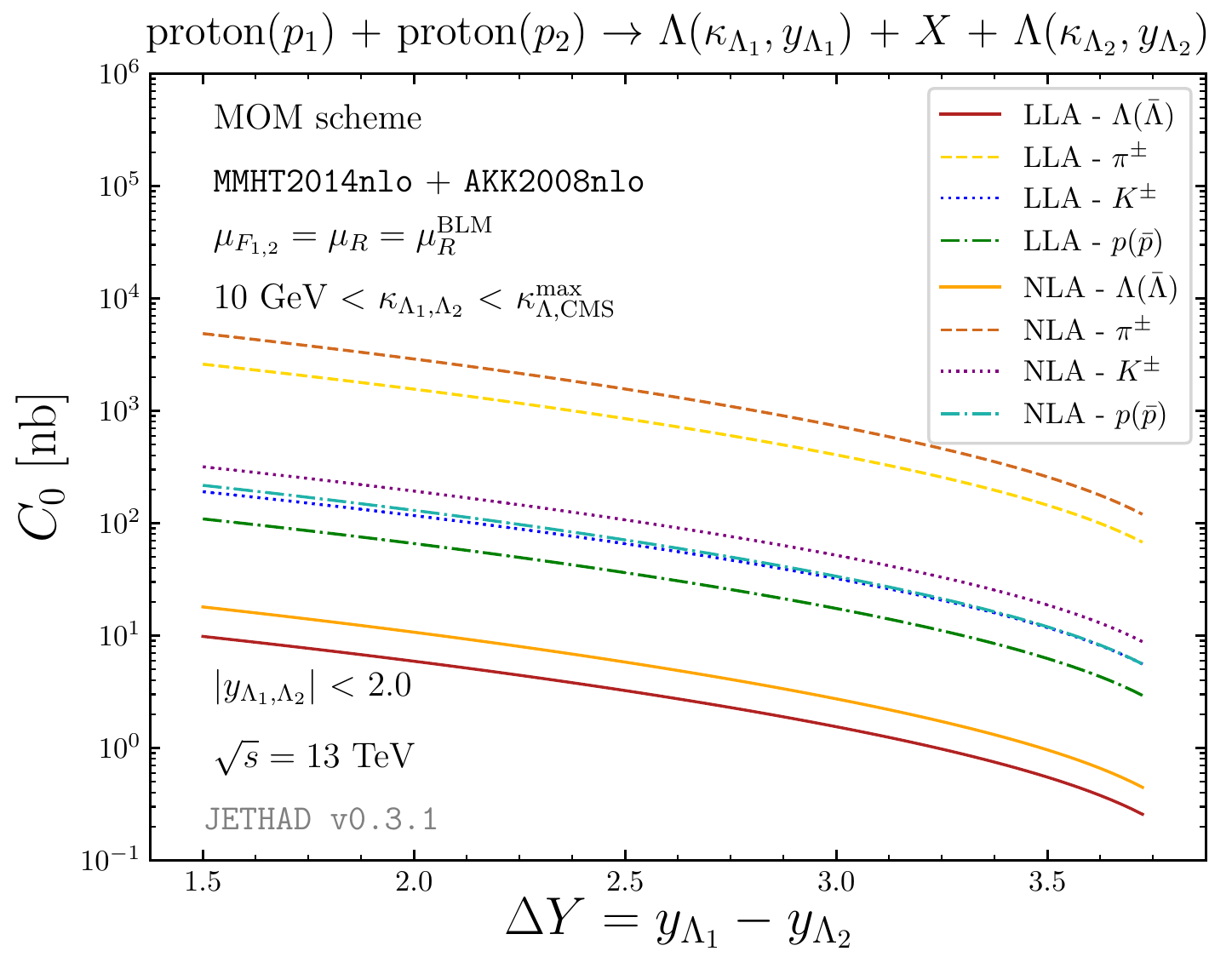}
 \\ \vspace{0.05cm}
 a) {\LL} channel

 \vspace{0.35cm}

 \includegraphics[scale=0.52,clip]{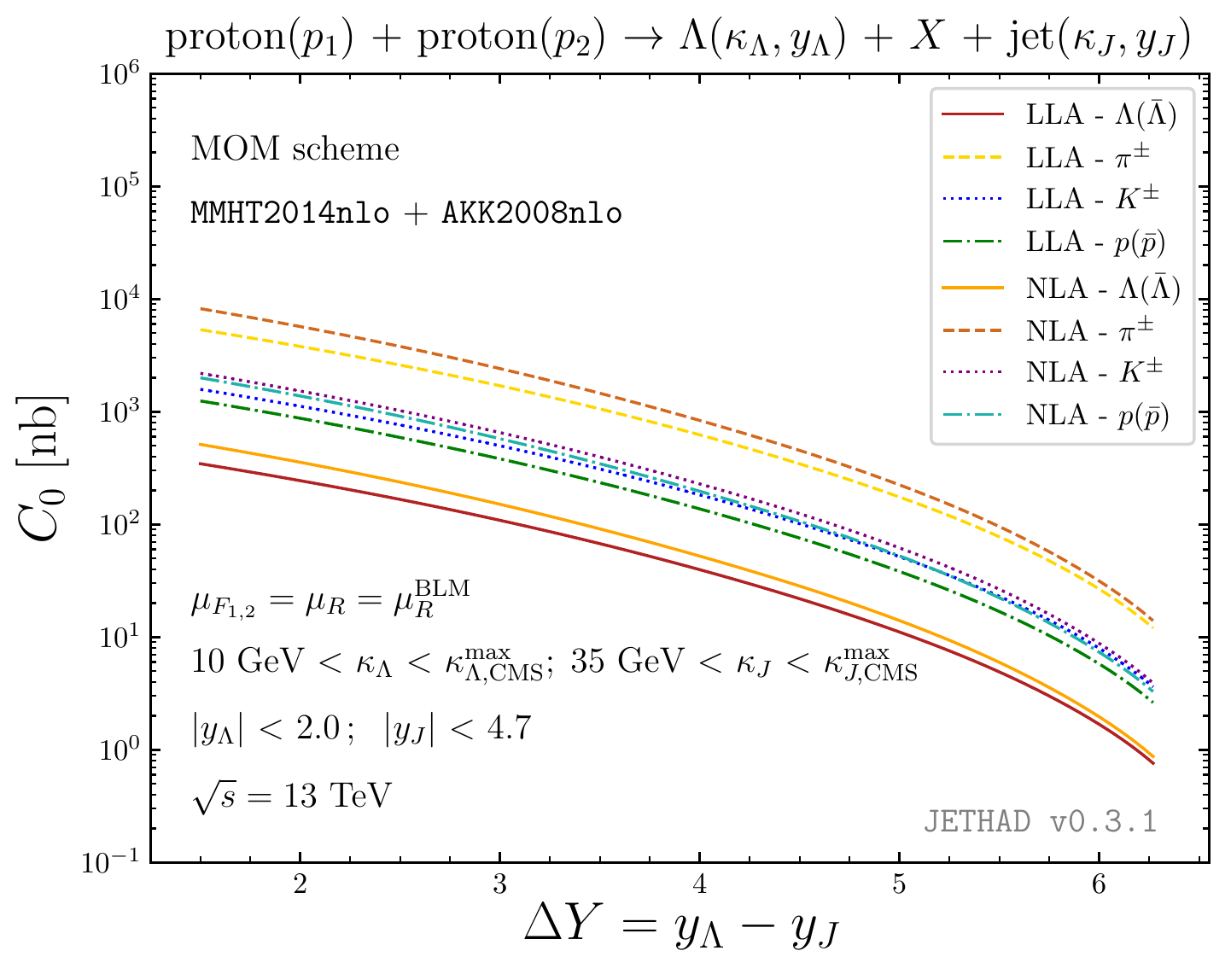}
 \hspace{0.25cm}
 \includegraphics[scale=0.52,clip]{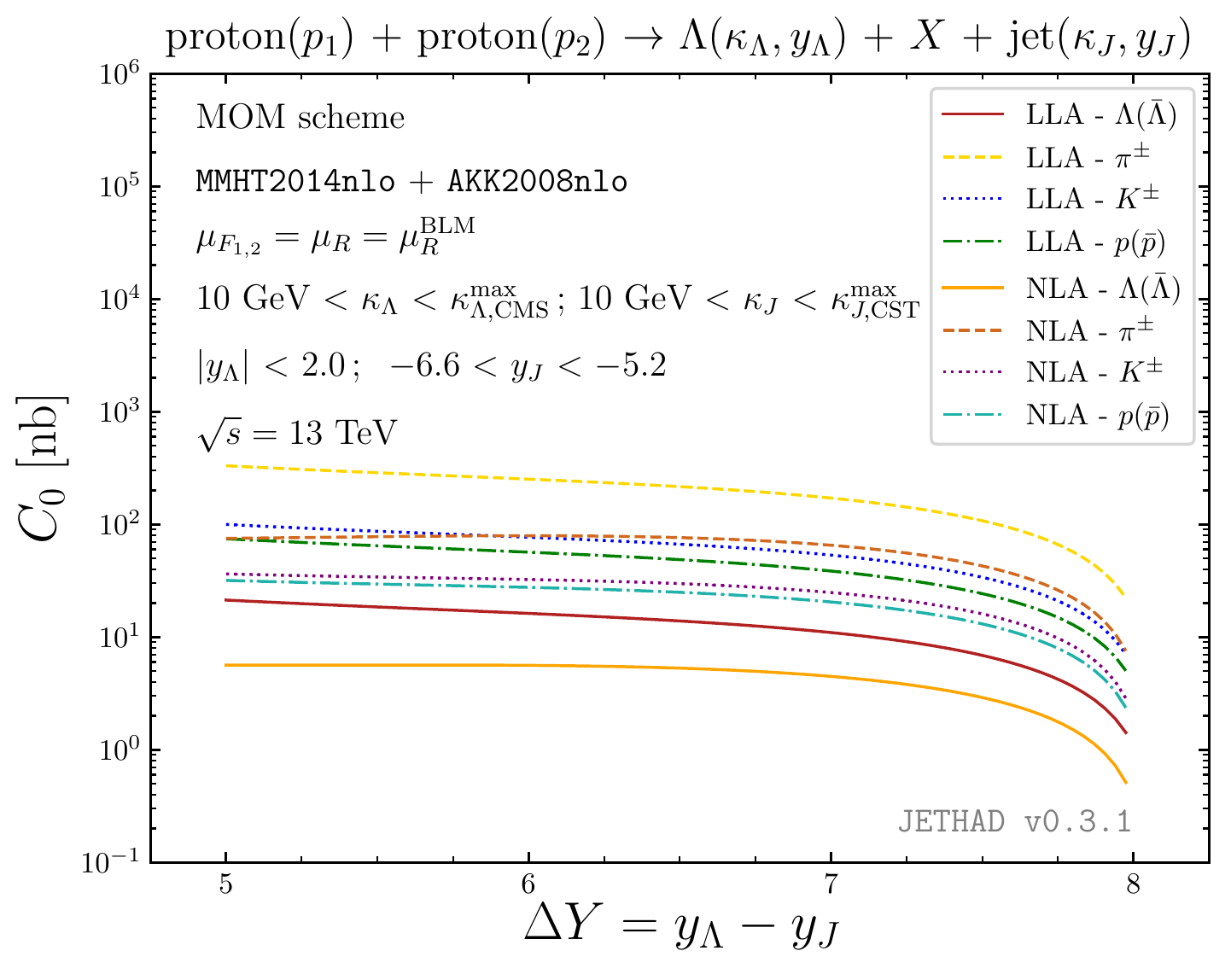}
 \\ \vspace{0.05cm}
 b) {\LJ} channel: \textit{CMS-jet} (left) and \textit{CASTOR-jet} (right) configurations

\caption{$\Delta Y$-dependence of the $\varphi$-averaged cross section, $C_0$, for the two considered final states (Fig.~\ref{fig:LL-LJ}) in the NLA BFKL accuracy and for $\sqrt{s} = 13$ TeV.}
\label{fig:C0-LL-LJ}
\end{figure}

Our predictions for the $\Delta Y$-dependence of $\varphi$-averaged cross sections, $C_0$, in all the considered production channels and kinematic configurations (Fig.~\ref{fig:C0-LL-LJ}) unambiguously state that the usual onset of the BFKL dynamics has come into play. Although the high-energy resummation leads to a rise with energy of the purely partonic cross section, the net effect of the convolution with PDFs (and FFs) is a downtrend with $\Delta Y$ of both LLA and NLA predictions. At the same time, next-to-leading corrections have opposite sign with respect to the leading ones, thus making NLA results constantly lower than
pure LLA ones. The main outcome of our results comes, however, from the comparison between $\Lambda$-hyperon emission(s) and lighter charged-hadron detections(s) -- pions, kaons or protons -- whose theoretical description exactly matches the setup of Section~\ref{cross_section}, the BFKL partonic cross section being convoluted with the respective hadron FFs as provided by the {\tt AKK2008nlo} sets. Cross sections in the {\LL}~(Fig.~\ref{fig:C0-LL-LJ}~a)) and in the {\LJ}~(Fig.~\ref{fig:C0-LL-LJ}~b)) channels are steadily lower with respect to the di-hadron and the hadron-jet ones, from one (pions and kaons) to three orders of magnitude (protons). This, together with the fact that the lower experimental cutoff for the $\Lambda$-particle identification is larger than the corresponding one for the light-hadron tagging (10 GeV versus 5 GeV, respectively), definitely validates our assertion on the opportunity to dampen, from the experimental point of view, minimum-bias contaminations. Considering $\Lambda$-hyperon emissions in the final states makes the comparison with data easier.

\begin{figure}[t]
\centering

   \includegraphics[scale=0.535,clip]{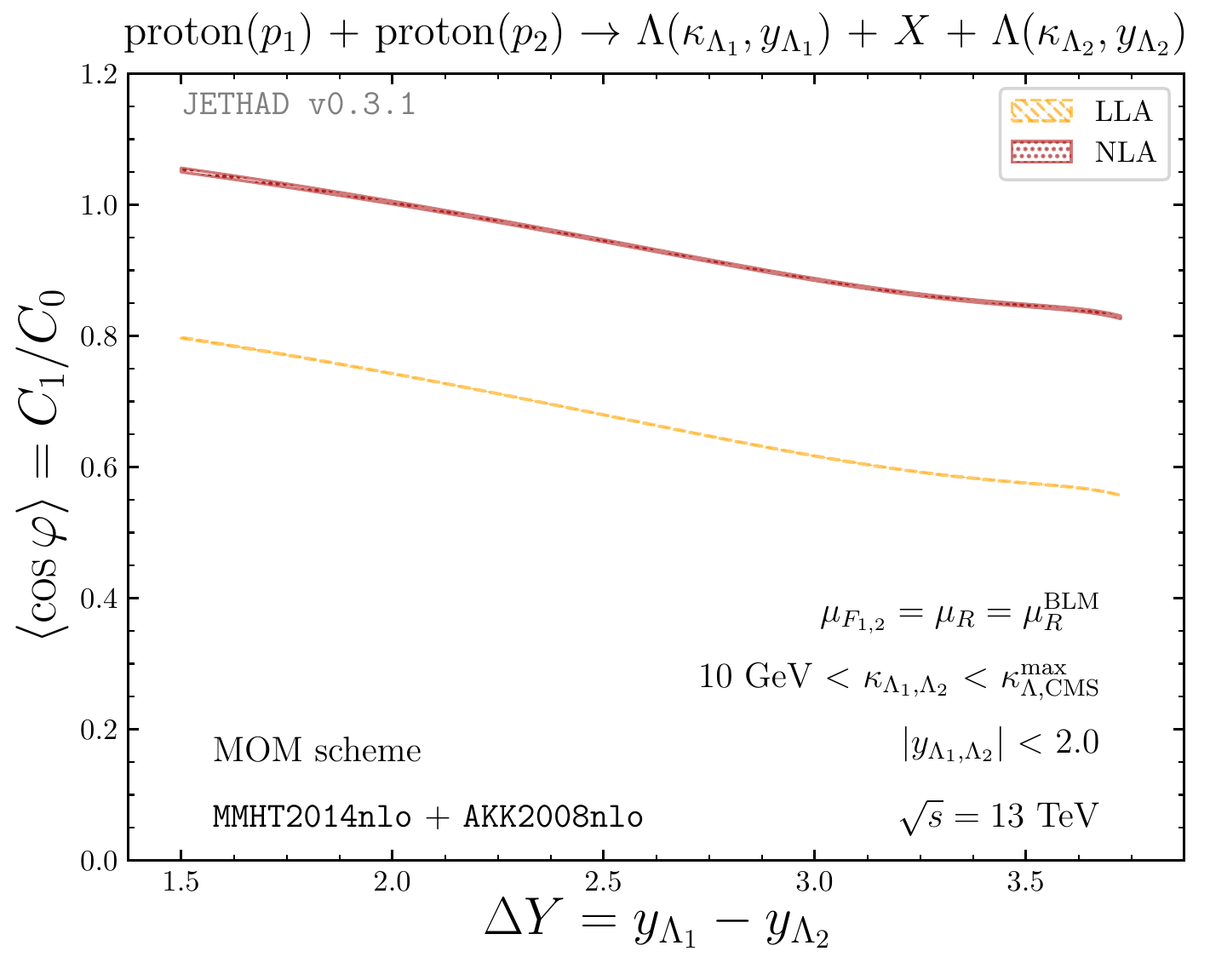}
   \hspace{0.25cm}
   \includegraphics[scale=0.535,clip]{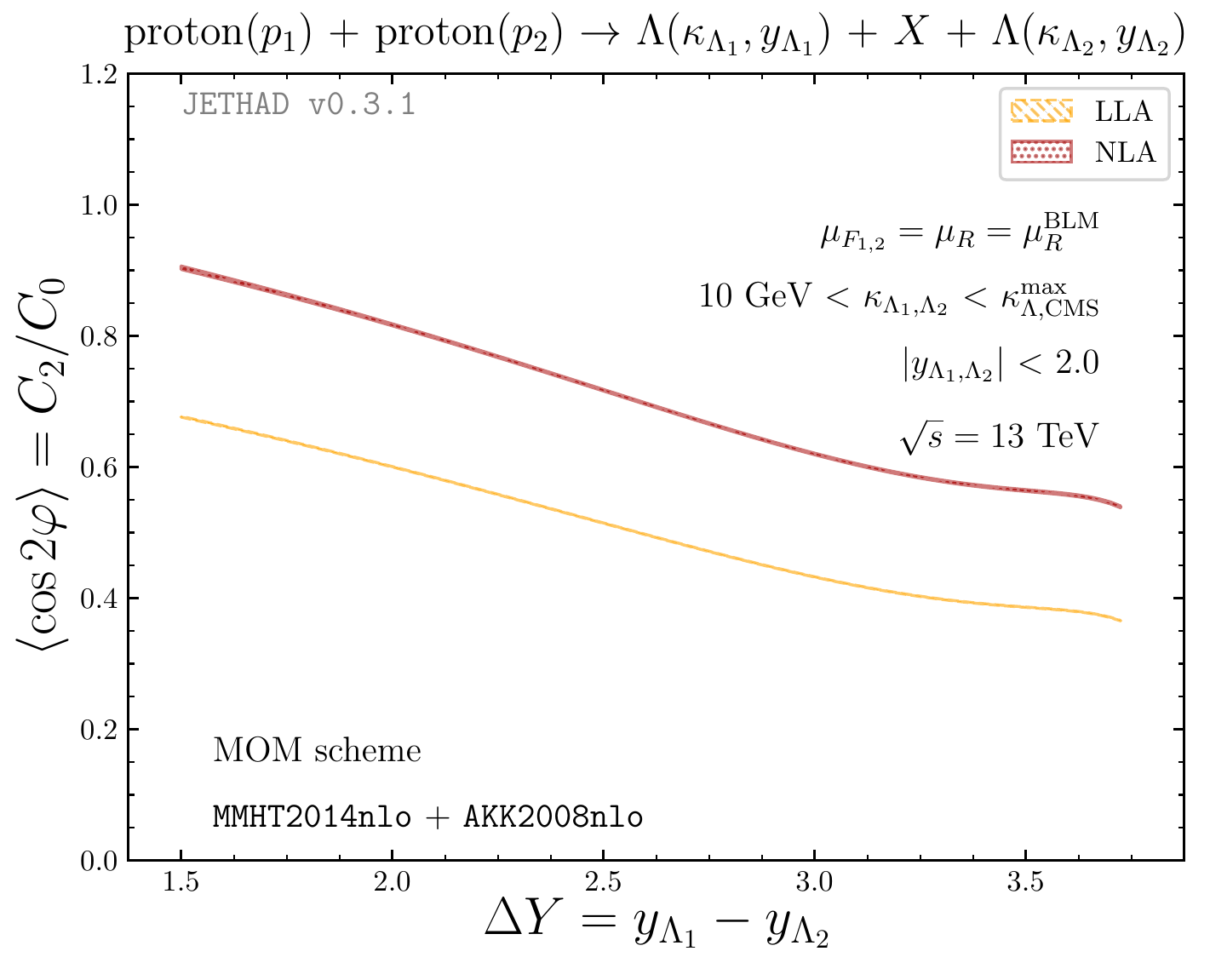}

   \includegraphics[scale=0.535,clip]{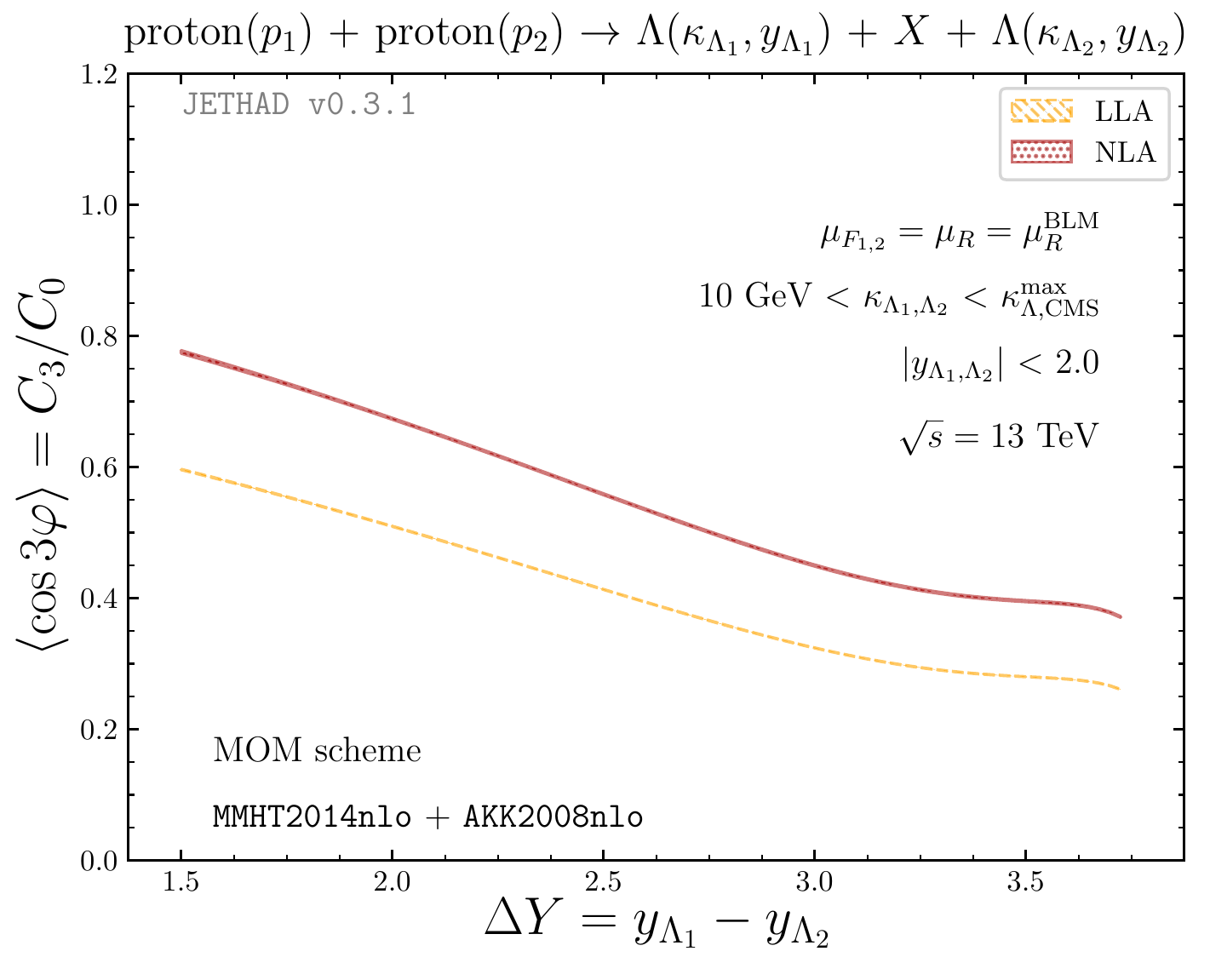}
   \hspace{0.25cm}
   \includegraphics[scale=0.535,clip]{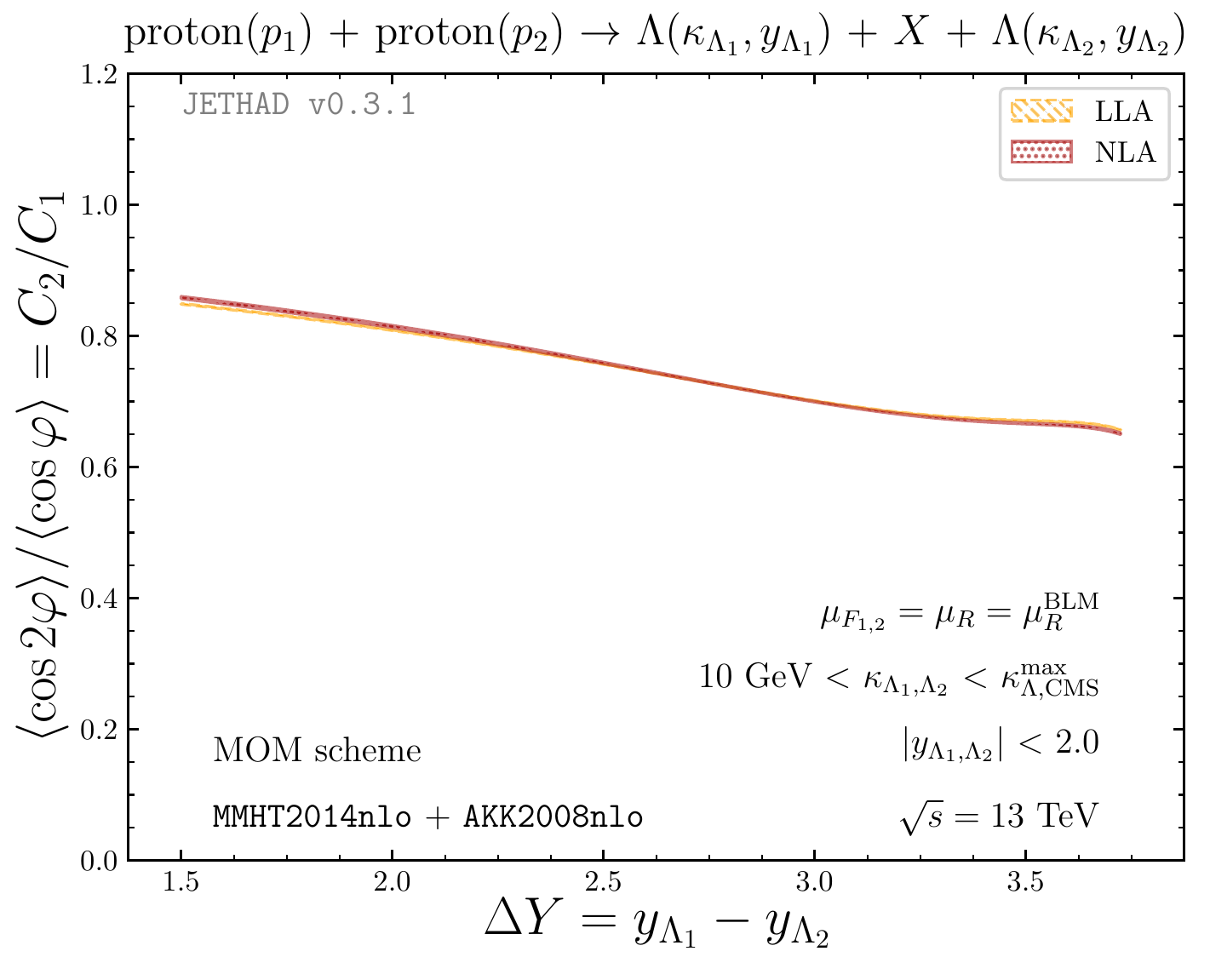}

\caption{$\Delta Y$-dependence of several azimuthal ratios, $R_{nm} \equiv C_n/C_m$, in the {\LL} channel (left panel of Fig.~\ref{fig:LL-LJ}) for
$\mu_{F1,2} = \mu_R = \mu_R^{\rm BLM}$ and $\sqrt{s} = 13$ TeV.}
\label{fig:LL-CMS}
\end{figure}

\begin{figure}[t]
\centering

   \includegraphics[scale=0.535,clip]{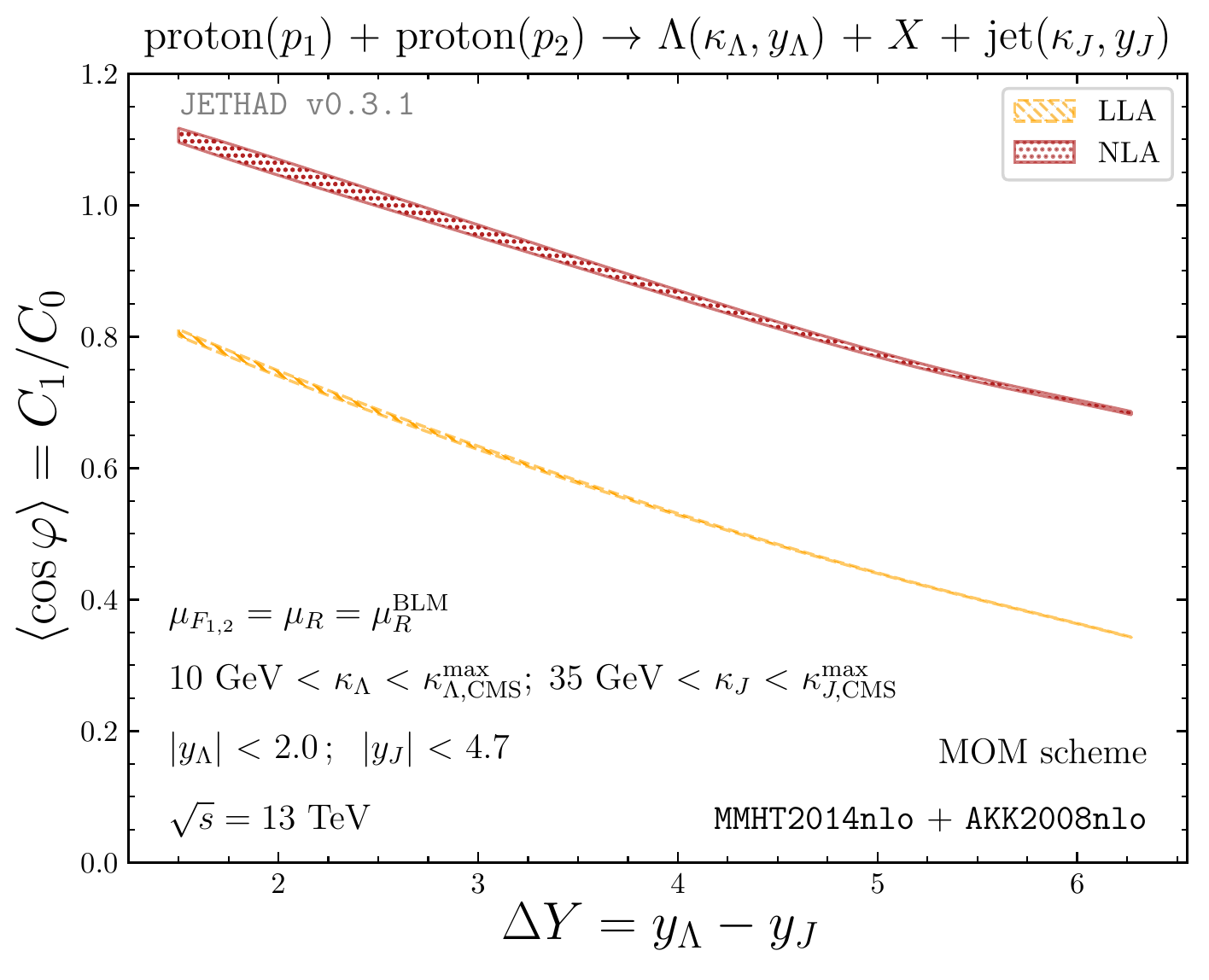}
   \hspace{0.25cm}
   \includegraphics[scale=0.535,clip]{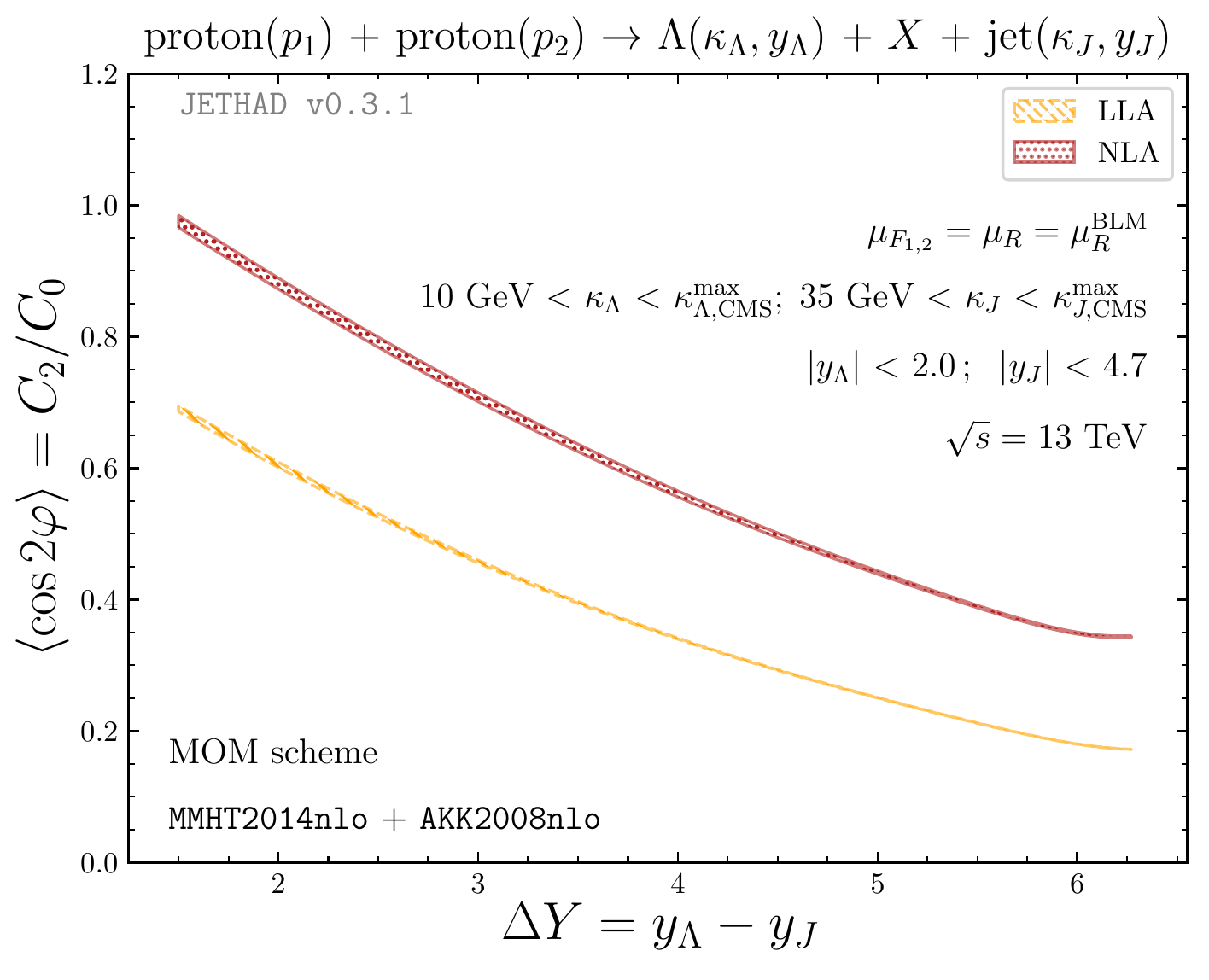}

   \includegraphics[scale=0.535,clip]{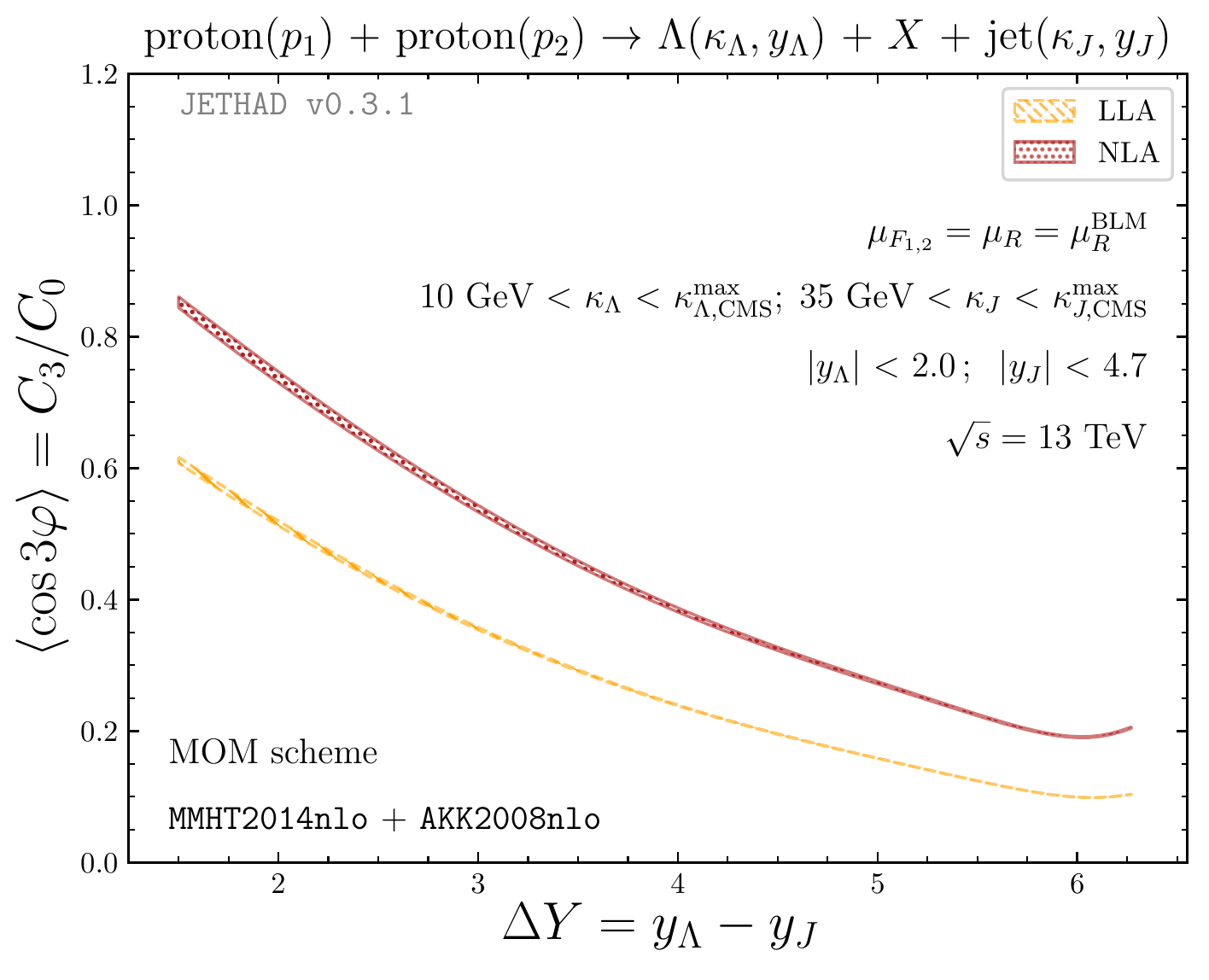}
   \hspace{0.25cm}
   \includegraphics[scale=0.535,clip]{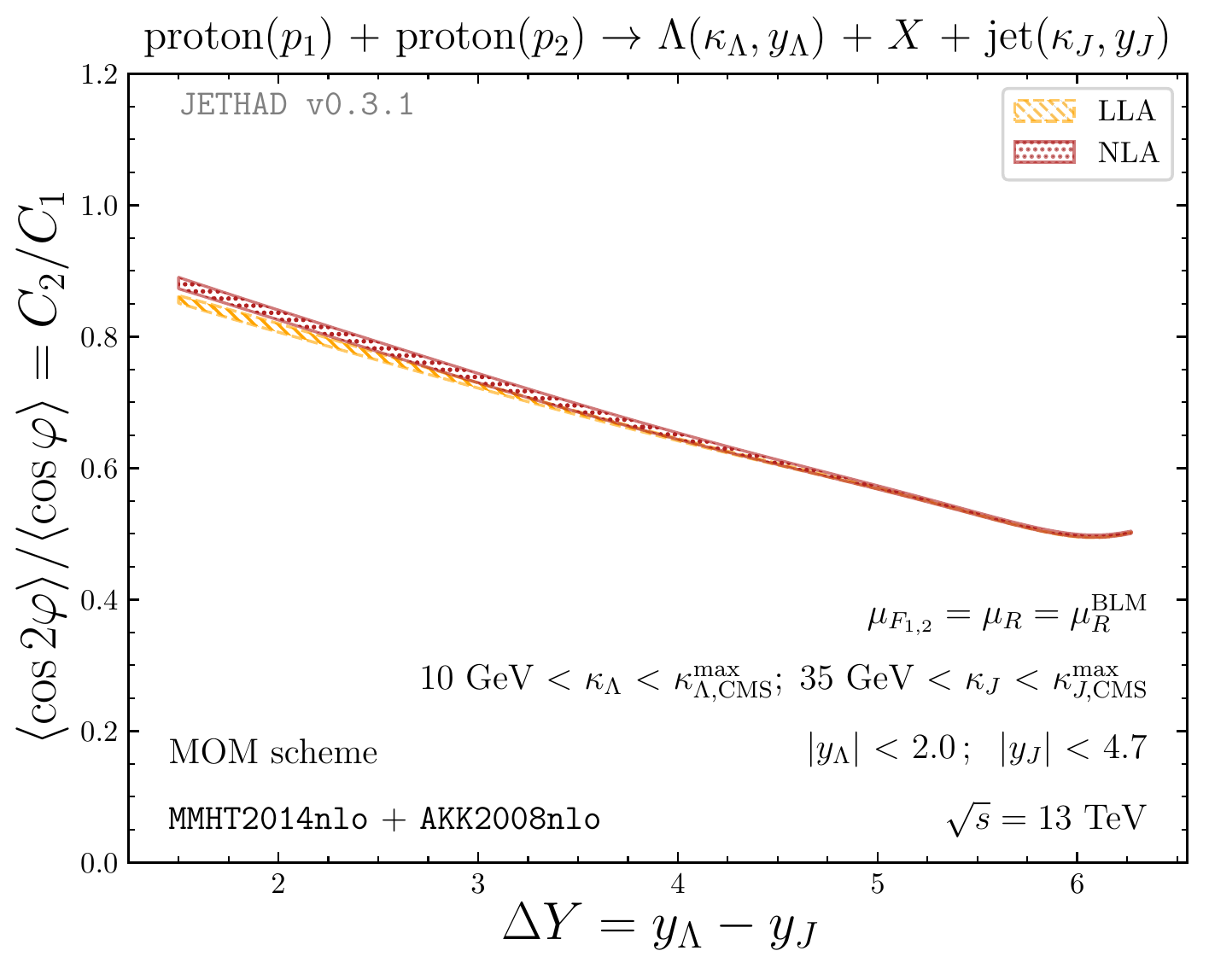}

\caption{$\Delta Y$-dependence of several azimuthal ratios, $R_{nm} \equiv C_n/C_m$, in the {\LJ} channel (right panel of Fig.~\ref{fig:LL-LJ}) for
$\mu_{F1,2} = \mu_R = \mu_R^{\rm BLM}$ and $\sqrt{s} = 13$ TeV (\textit{CMS-jet} configuration).}
\label{fig:LJ-CMS}
\end{figure}

\begin{figure}[t]
\centering

   \includegraphics[scale=0.535,clip]{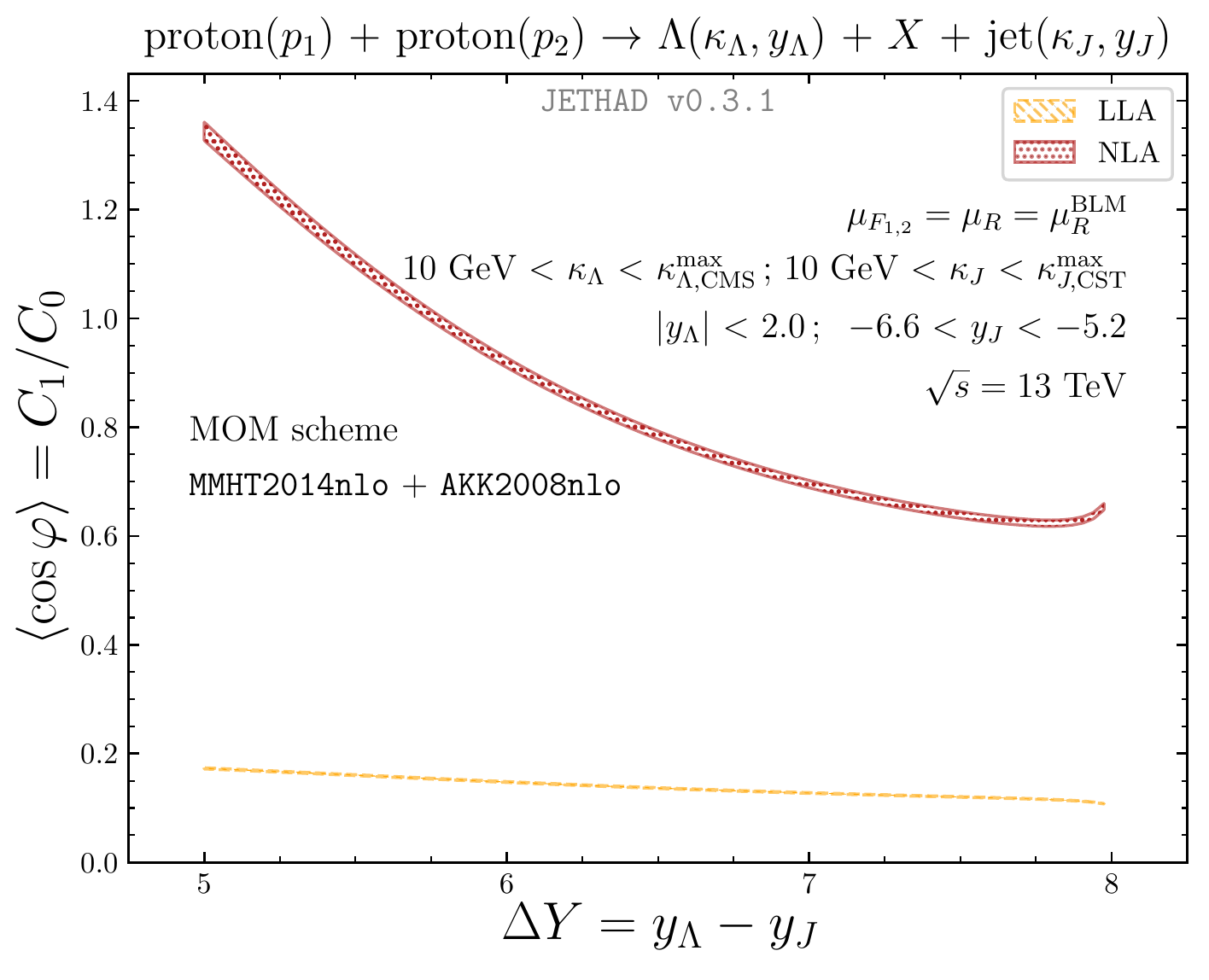}
   \hspace{0.25cm}
   \includegraphics[scale=0.535,clip]{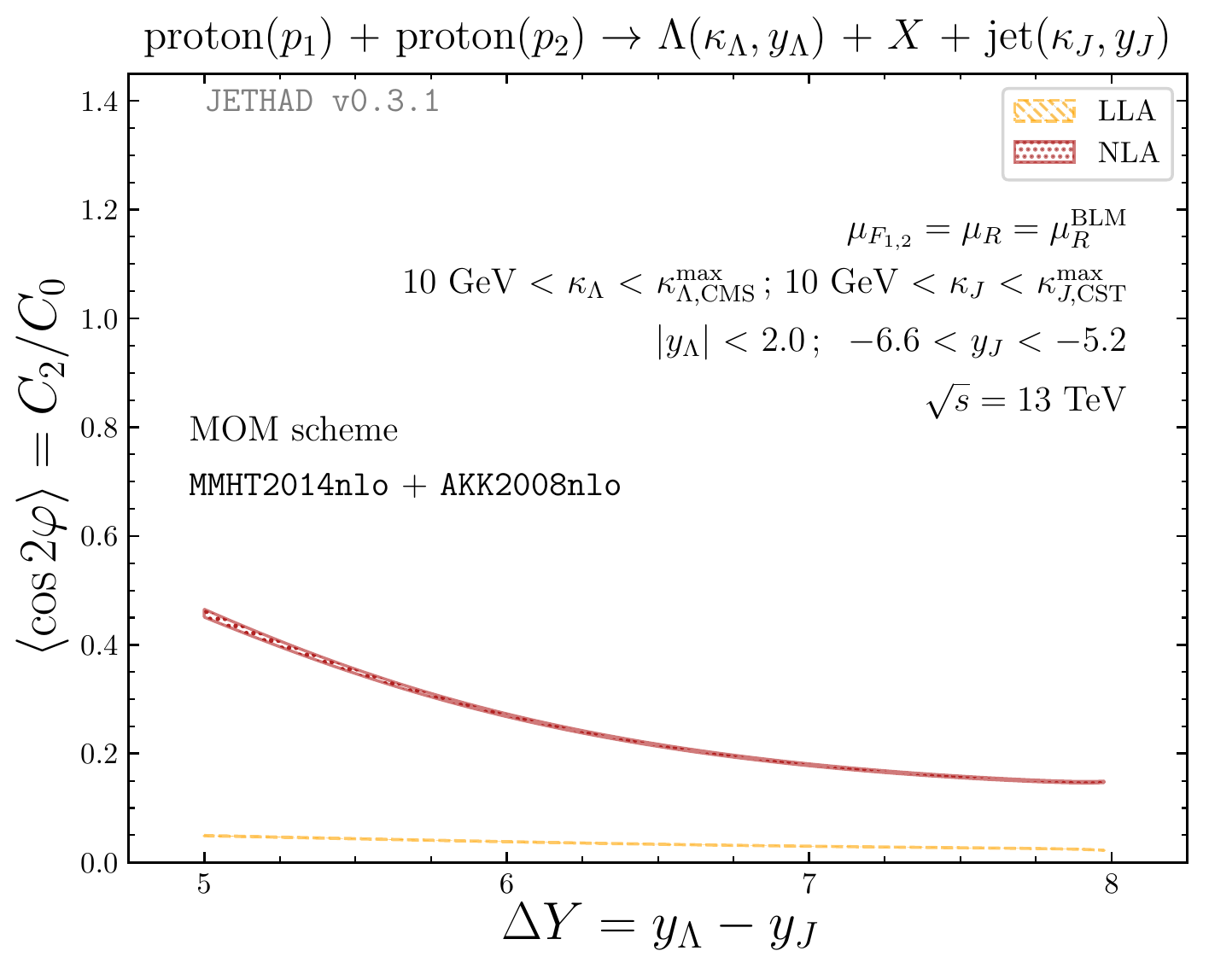}

   \includegraphics[scale=0.535,clip]{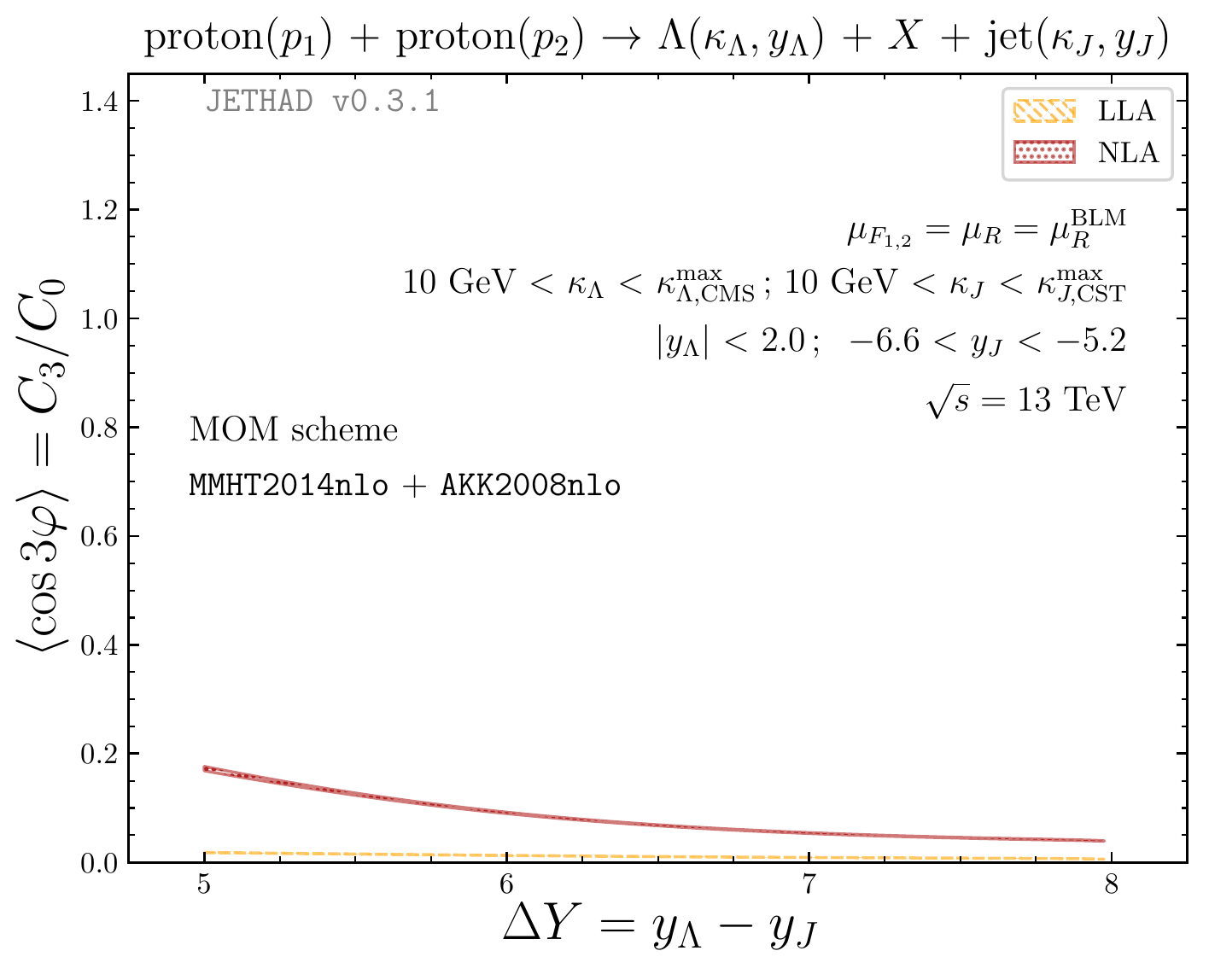}
   \hspace{0.25cm}
   \includegraphics[scale=0.535,clip]{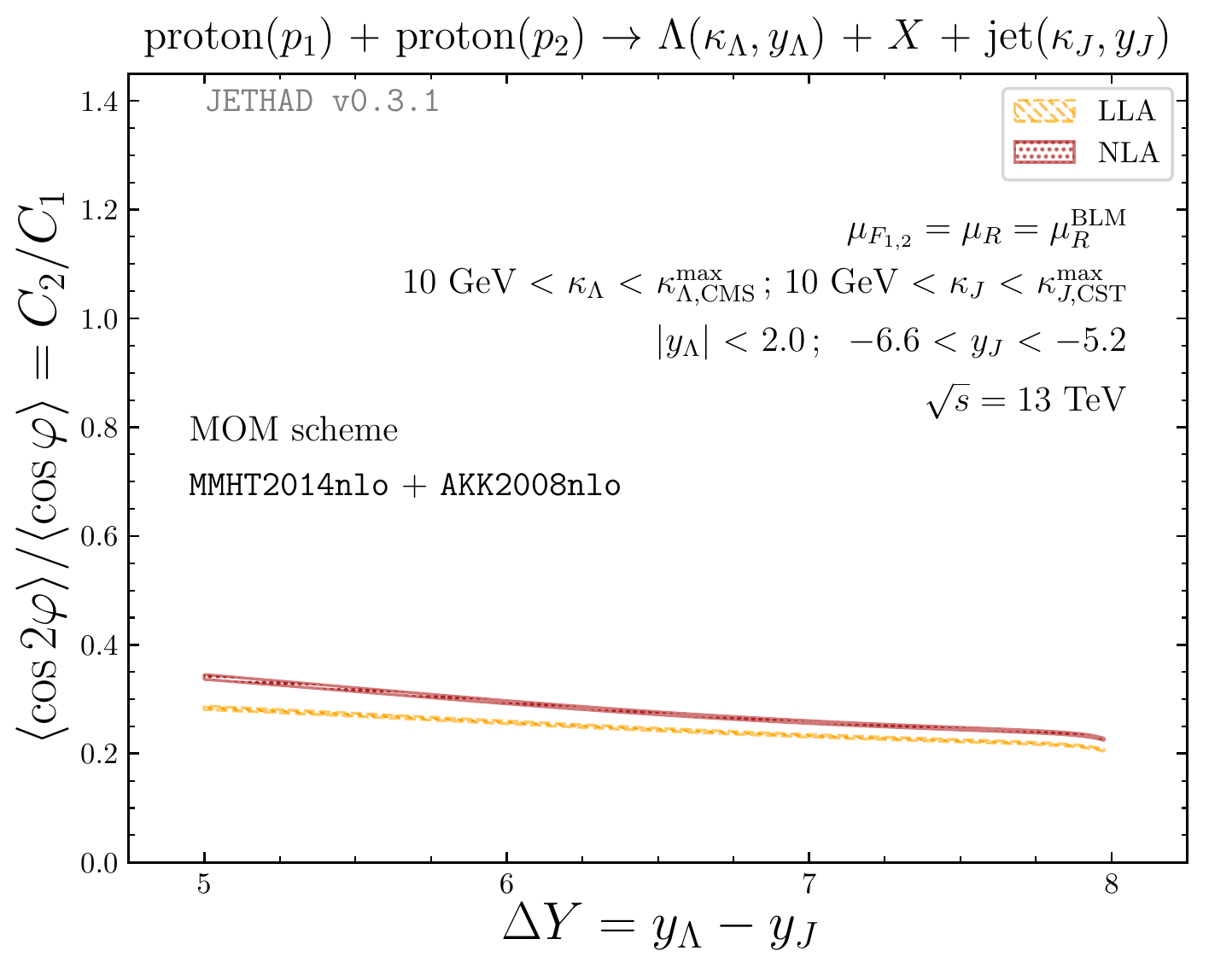}

\caption{$\Delta Y$-dependence of several azimuthal ratios, $R_{nm} \equiv C_n/C_m$, in the {\LJ} channel (right panel of Fig.~\ref{fig:LL-LJ}) for
$\mu_{F1,2} = \mu_R = \mu_R^{\rm BLM}$ and $\sqrt{s} = 13$ TeV (\textit{CASTOR-jet} configuration).}
\label{fig:LJ-CST}
\end{figure}

In the next figures we present the $\Delta Y$-dependence of some azimuthal-correlation moments, $R_{nm} \equiv C_n/C_m$, in the {\LL}~channel~(Fig.~\ref{fig:LL-CMS}) and in the {\LJ} one, for both the \textit{CMS-jet}~(Fig.~\ref{fig:LJ-CMS}) and the \textit{CASTOR-jet} (Fig.~\ref{fig:LJ-CST}) final-state ranges. Here, the emission of undetected gluons (the-$X$ subsystem in Eq.~(\ref{process})) increases with the final-state rapidity interval, $\Delta Y$, thus leading to the falloff with $\Delta Y$ of all the azimuthal correlations. Next-to-leading corrections are responsible for a ``recorrelation'' the $R_{nm}$ ratios, thus making the NLA results larger than the pure-LLA ones. In all the considered figures, the value of $R_{10}$ exceeds one in the case of small rapidity distance, an unphysical effect particularly relevant in the \textit{CASTOR-jet} configuration, which deserves further attention. We provide with an explanation of this issue in the last part of this Section.

Having recovered the usual high-energy trend for the azimuthal correlations corroborates the validity of our theoretical approach, suited to the description of lighter charged hadrons, its validity being evidently holding also in the analysis of $\Lambda$ hyperons. 

\begin{figure}[t]
\centering

   \includegraphics[scale=0.535,clip]{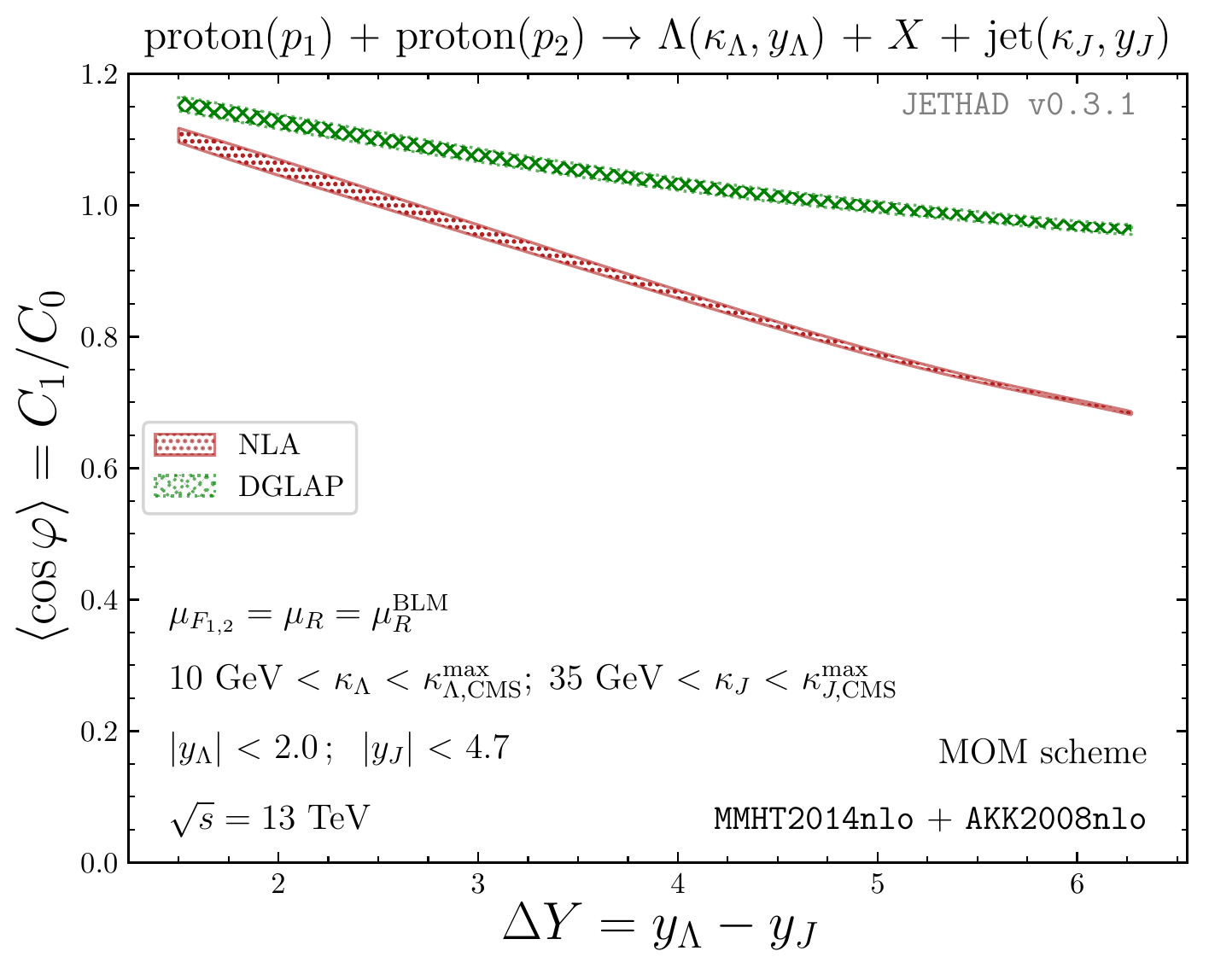}
   \hspace{0.25cm}
   \includegraphics[scale=0.535,clip]{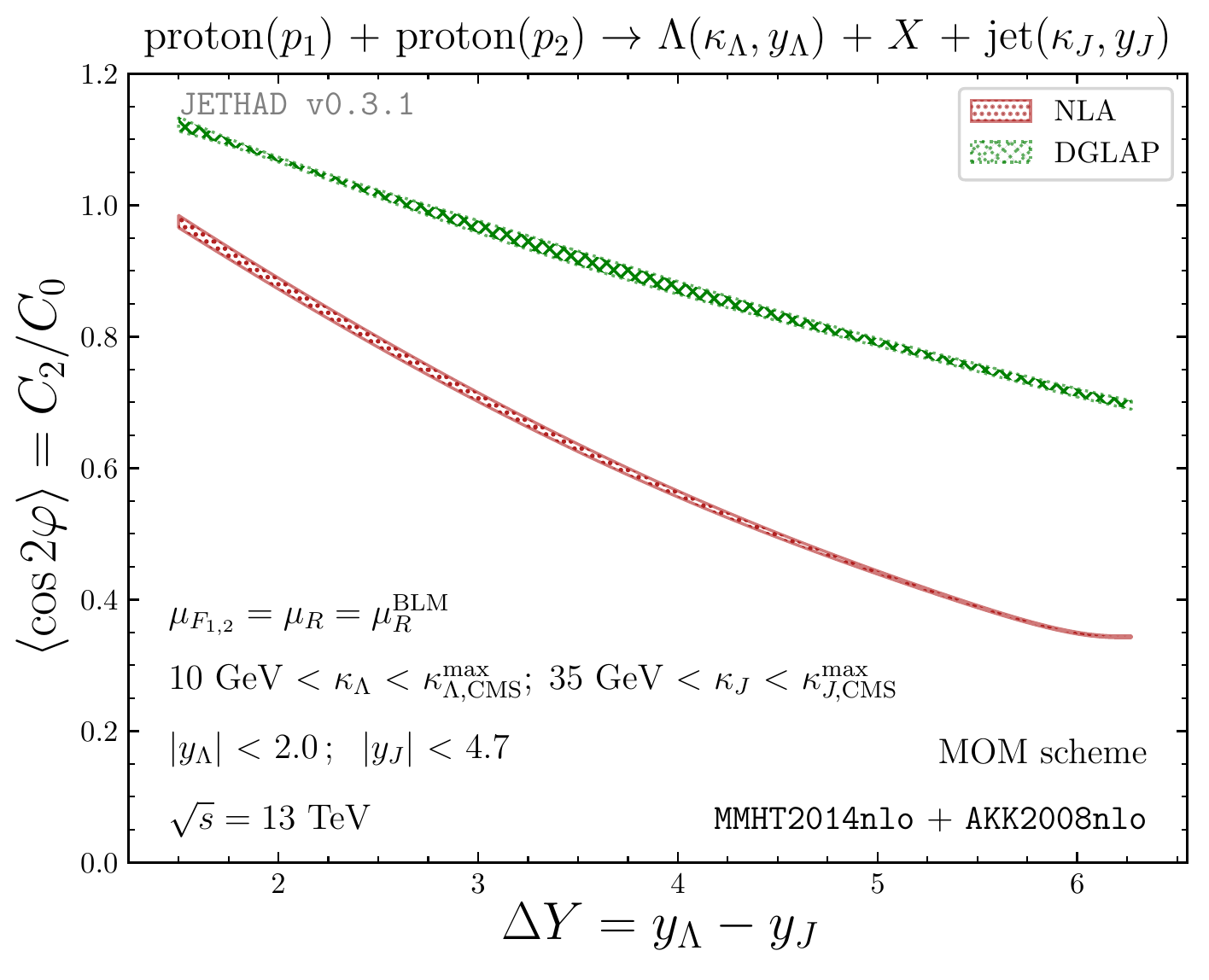}
 \\ \vspace{0.05cm}
 a) {\LJ} channel: \textit{CMS-jet} configuration

 \vspace{0.35cm}

   \includegraphics[scale=0.535,clip]{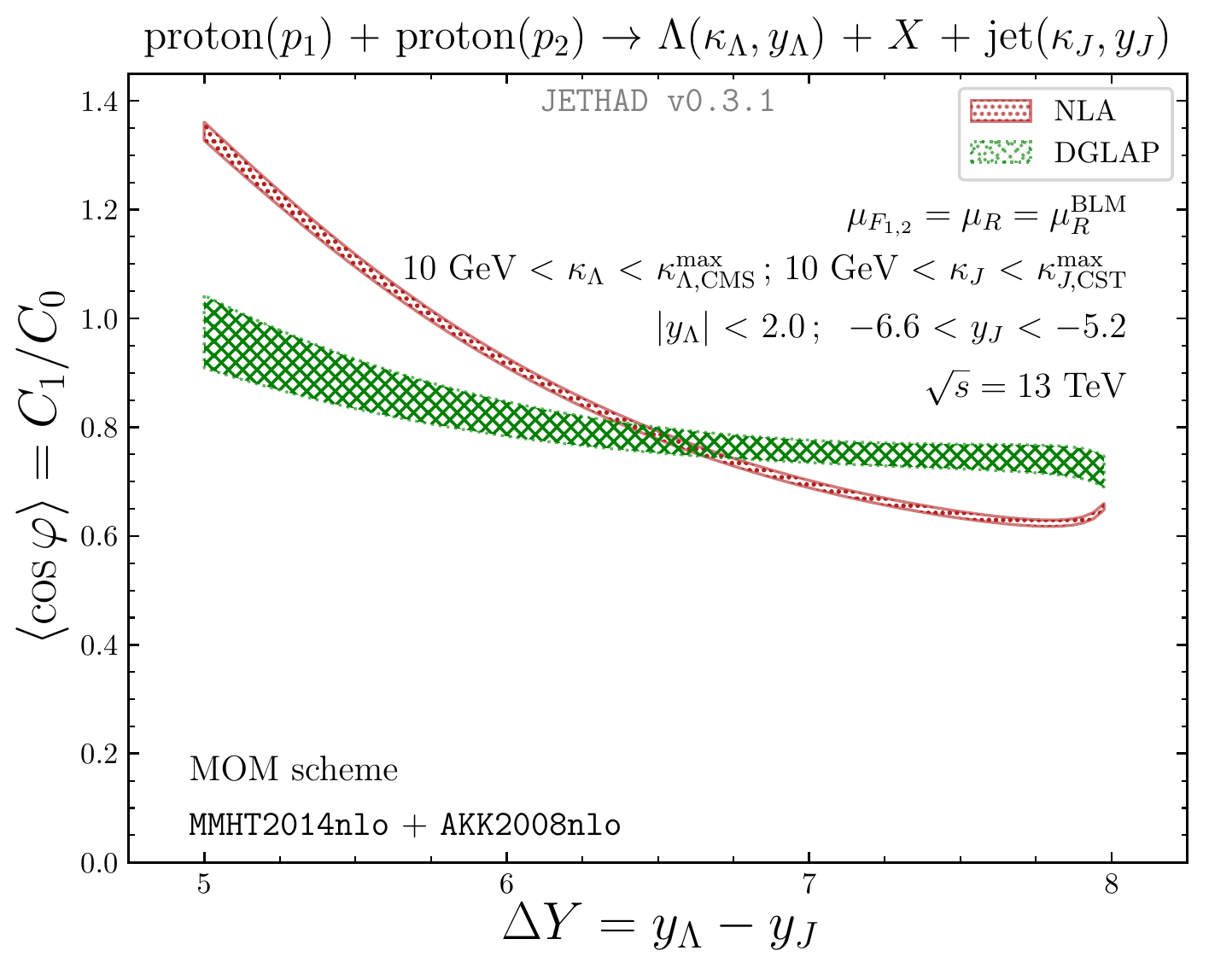}
   \hspace{0.25cm}
   \includegraphics[scale=0.535,clip]{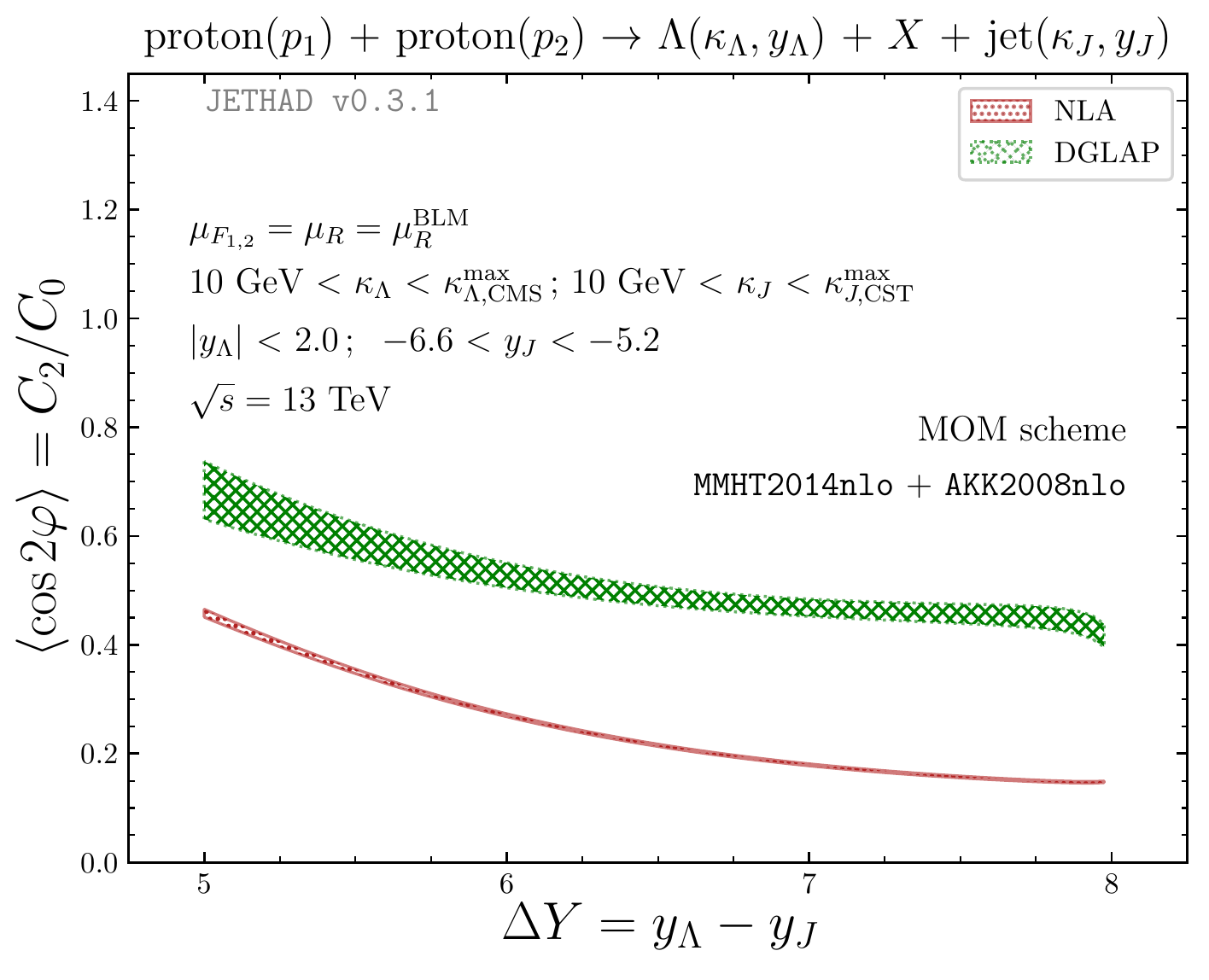}
 \\ \vspace{0.05cm}
 b) {\LJ} channel: \textit{CASTOR-jet} configuration

\caption{Comparison of BFKL and high-energy DGLAP predictions for the two azimuthal ratios, $R_{10}$ and $R_{20}$, as functions of the rapidity interval, $\Delta Y$, in the {\LJ} channel and for $\mu_{F1,2} = \mu_R = \mu_R^{\rm BLM}$ and $\sqrt{s} = 13$ TeV. Upper (lower) panels refer to the \textit{CMS-jet} (\textit{CASTOR-jet}) event selection.}
\label{fig:LJ-CMS-CST-BvD}
\end{figure}

Finally, in Fig.~\ref{fig:LJ-CMS-CST-BvD} we compare our BFKL predictions for $R_{10}$ and $R_{20}$ in the {\LJ}~channel with the corresponding ones obtained in the high-energy limit of DGLAP. Panels a) of Fig.~\ref{fig:LJ-CMS-CST-BvD} show the $\Delta Y$-behavior of the two azimuthal ratios in the \textit{CMS-jet} configuration, whereas the \textit{CASTOR-jet} event-selection case is presented in panels b).
As expected, a net distance between BFKL and DGLAP emerges and becomes more and more evident as the rapidity distance, $\Delta Y$, grows. At variance with BFKL, in the DGLAP case only a limited number of gluons, fixed by the truncation order of the perturbative series, can be inclusively emitted. The choice of two distinct final-state objects, namely a $\Lambda$ hyperon and a jet, naturally translates in an asymmetric selection for the transverse-momentum ranges. This quenches the Born contribution and enhances the discrepancy between the two approaches.
The overall outcome of this dedicated BFKL-vs-DGLAP analysis in the {\LJ}~channel is in line with patterns found in the Mueller--Navelet channel~\cite{Celiberto:2015yba} as well as in the case of inclusive light-charged hadron detection accompanied by a jet emission~\cite{Celiberto:2020wpk}.

\begin{figure}[bt]
\centering

   \includegraphics[scale=0.535,clip]{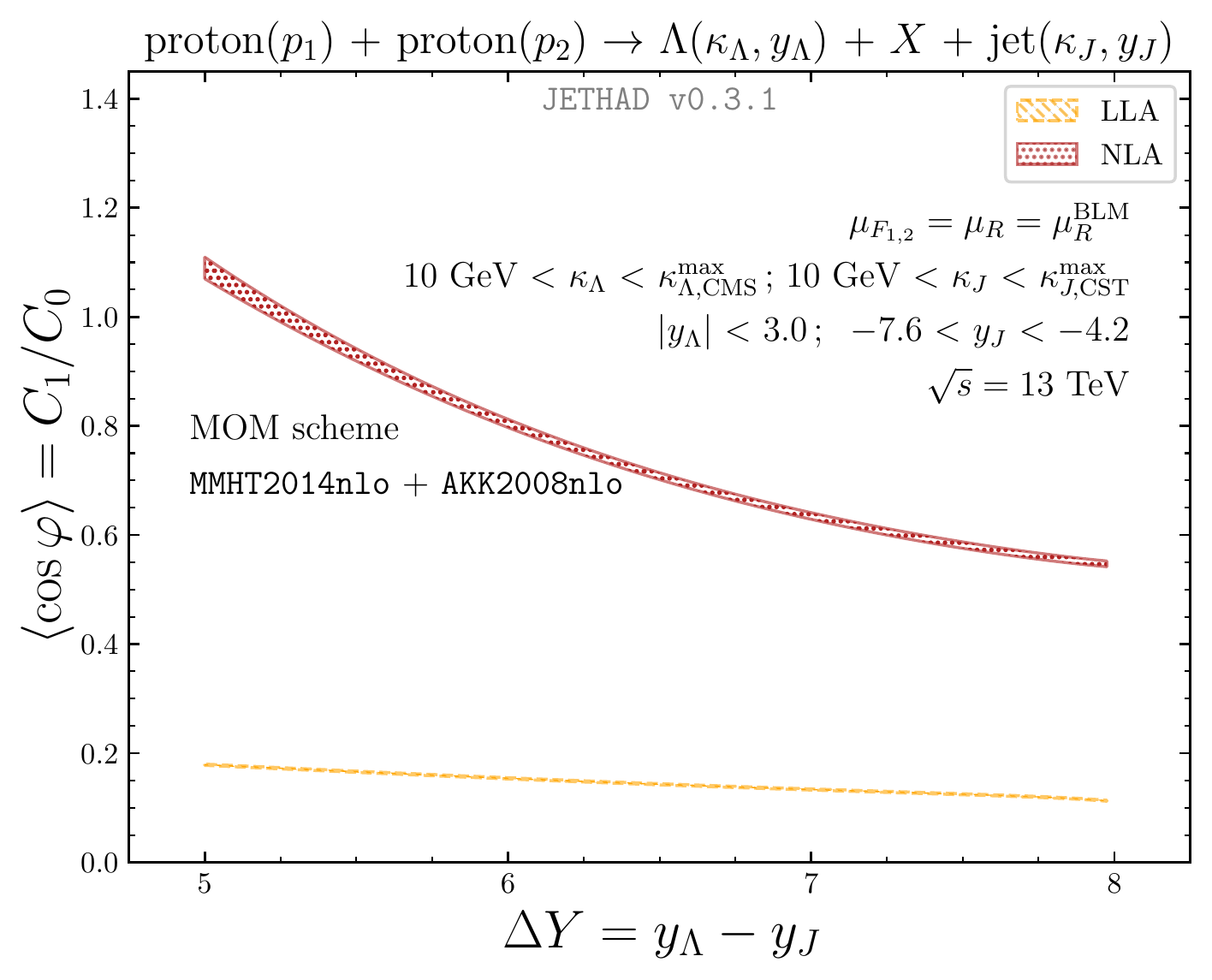}
   \hspace{0.25cm}
   \includegraphics[scale=0.535,clip]{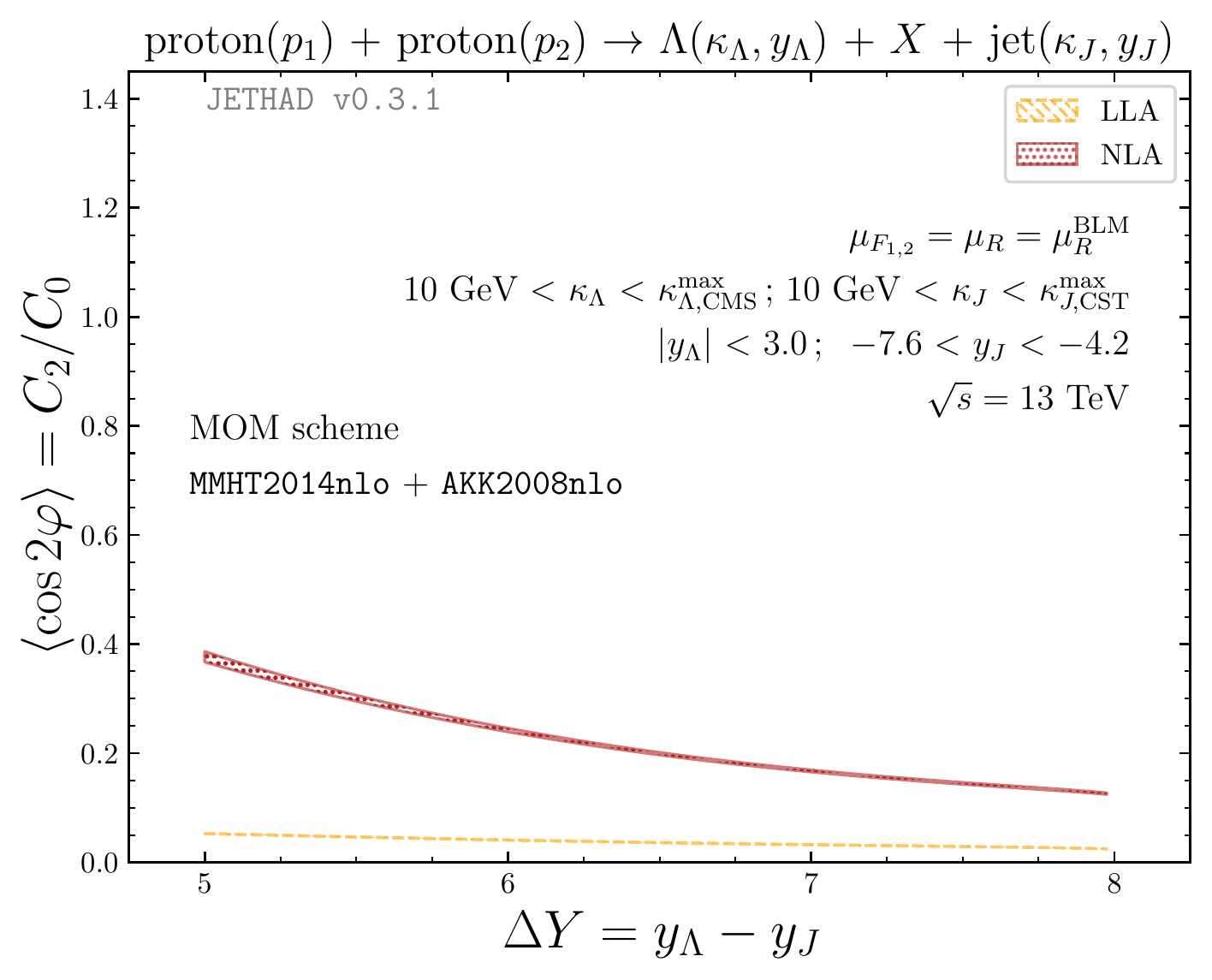}

   \includegraphics[scale=0.535,clip]{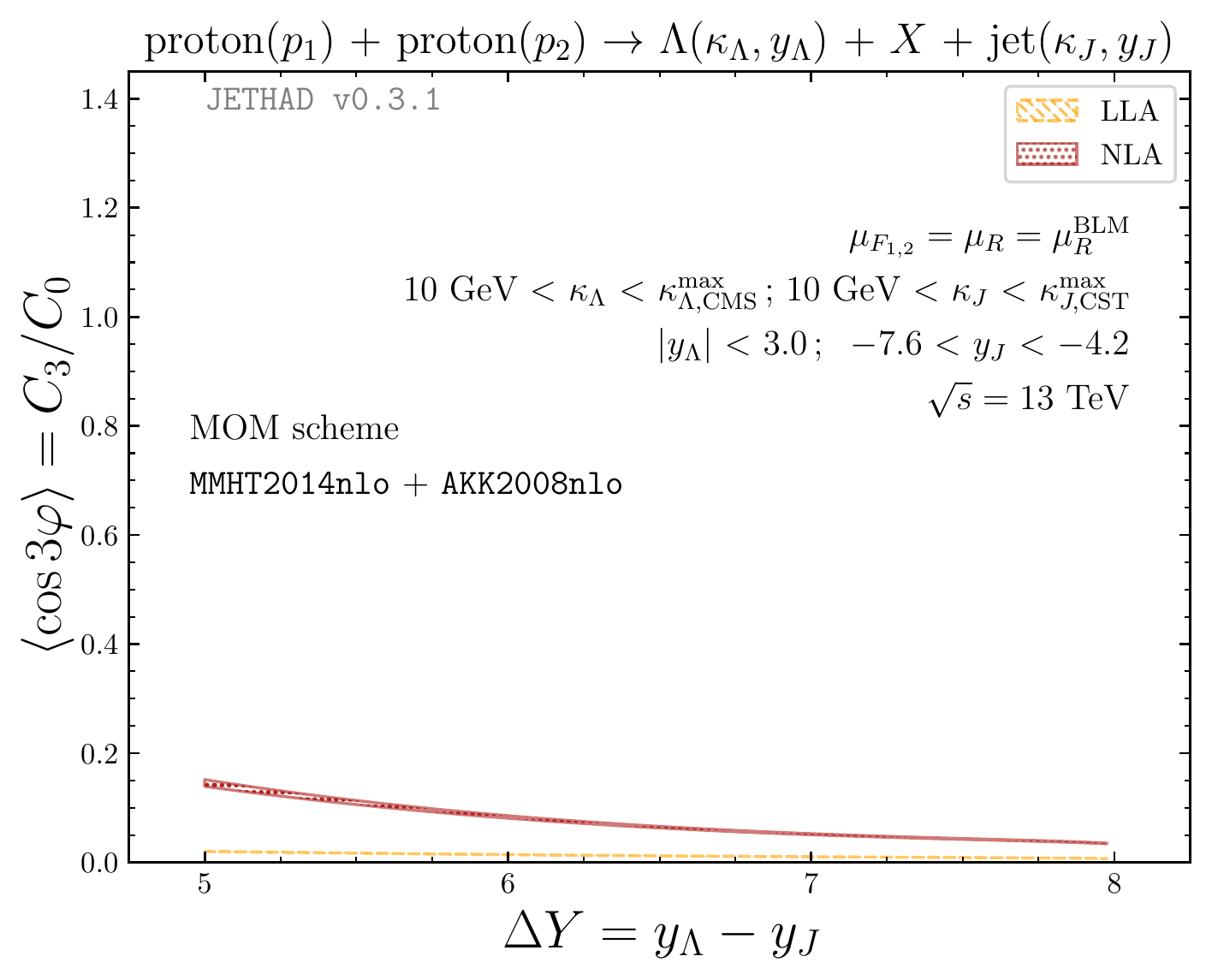}
   \hspace{0.25cm}
   \includegraphics[scale=0.535,clip]{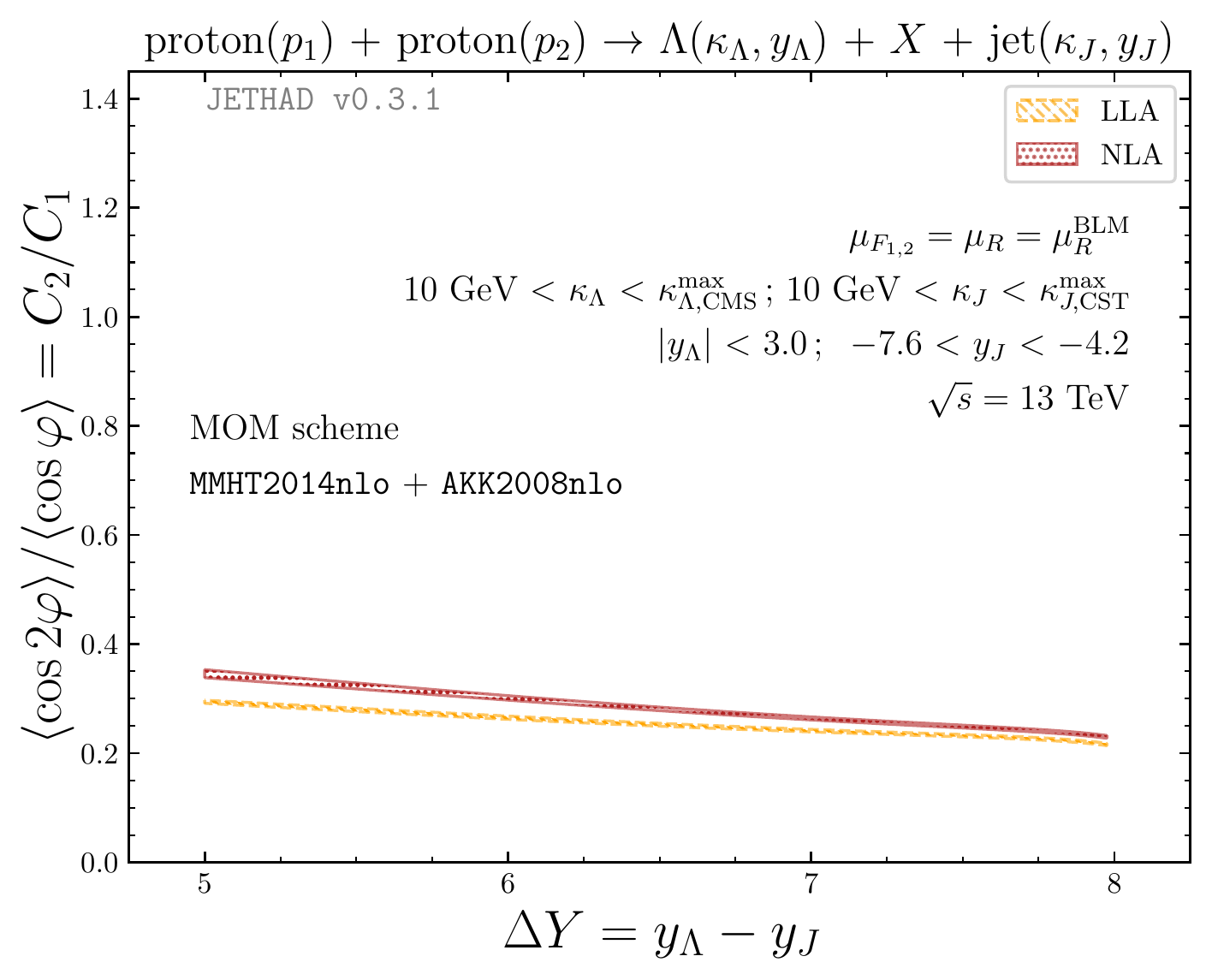}

\caption{$\Delta Y$-dependence of several azimuthal ratios, $R_{nm} \equiv C_n/C_m$, in the {\LJ} channel for
$\mu_{F1,2} = \mu_R = \mu_R^{\rm BLM}$ and $\sqrt{s} = 13$ TeV (\textit{CASTOR-jet} configuration with enlarged rapidity ranges).}
\label{fig:LJ-CST-lyb}
\end{figure}

\begin{figure}[t]
\centering

   \includegraphics[scale=0.535,clip]{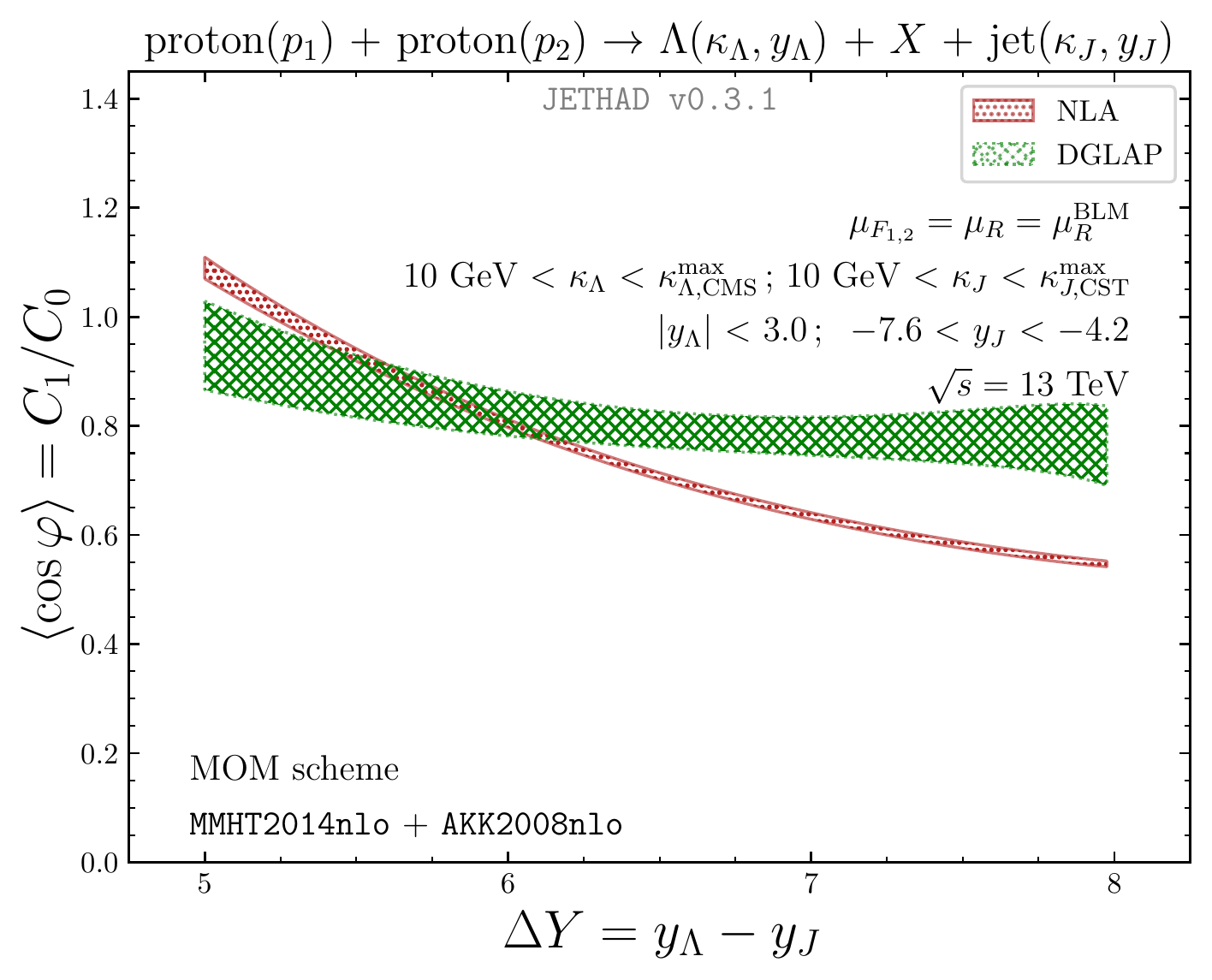}
   \hspace{0.25cm}
   \includegraphics[scale=0.535,clip]{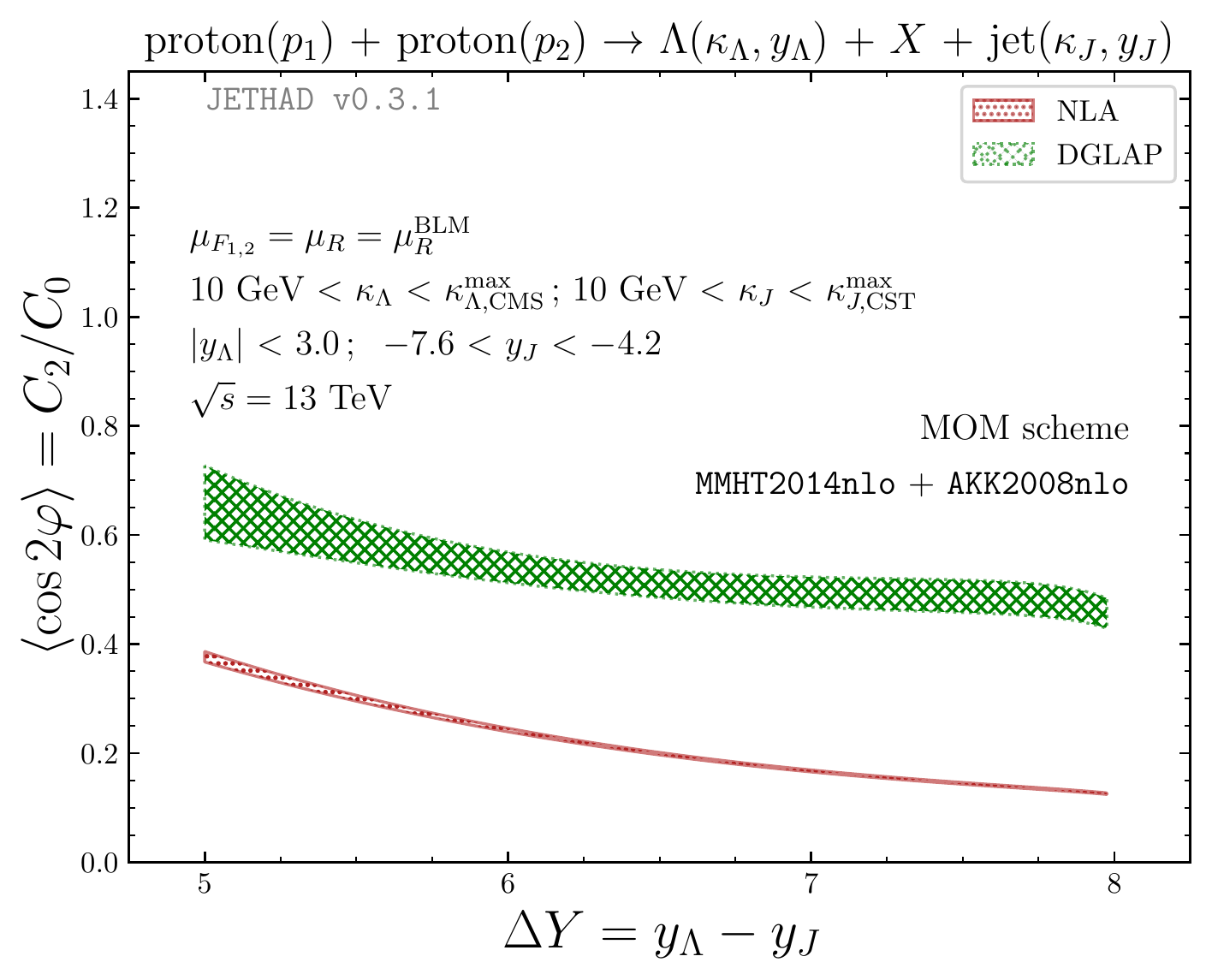}


\caption{Comparison of BFKL and high-energy DGLAP predictions for the two azimuthal ratios, $R_{10}$ and $R_{20}$, as functions of the rapidity interval, $\Delta Y$, in the {\LJ} channel and for $\mu_{F1,2} = \mu_R = \mu_R^{\rm BLM}$ and $\sqrt{s} = 13$ TeV (\textit{CASTOR-jet} configuration with enlarged rapidity ranges).}
\label{fig:LJ-CST-lyb-BvD}
\end{figure}

Now we come back to the issue about the unphysical values of $R_{10}$ in the small-$\Delta Y$ region.
In the \textit{CASTOR-jet} case, the allowed rapidity 
band for the detected jet is close to its kinematic boundary, that corresponds 
to the longitudinal momentum fraction of the jet parent parton going to its 
limit, $x=1$. For instance, for $\kappa_J = 10 \div 15$ GeV and $y_J=6.6$, the value of the parton $x$ is in the range, 0.57 $\div$ 0.85, at LO. In this kinematics 
the undetected gluon radiation from large-$x$ partons gets effectively restricted.
In general such kinematical restriction leads, in inclusive observables, to an
incomplete cancellation between virtual and real gluon-emission contributions,
which results in the appearance of large Sudakov-type double logarithms (threshold
double logarithms) in the perturbative series, that have to be resummed to all orders
to obtain valuable predictions.
In the BFKL approach we systematically resum contributions of leading (LLA) and 
first non-leading (NLA) logarithms of energy, whereas effects of the above mentioned
threshold double logarithms are implicitly accounted for only through the NLO
corrections to the jet impact factor. Note also that, at the LLA BFKL level, one
is completely insensitive to these threshold double logarithms. 
In technical terms, inspection of our result for NLO jet impact factors shows
that it contains some terms which include the plus prescription, like
$(\ln(1-\xi)/(1-\xi))_+$. After its convolution with the quark PDF, we obtain
an effect roughly proportional to the quark PDF derivative. Therefore, for
\textit{CASTOR-jet} large-$x$ kinematics, one gets a large NLO correction to the jet impact factor, since the PDF derivative value is much bigger than the value of the PDF
at large $x$. NLA results for the \textit{CASTOR-jet} case shown in Figs.~\ref{fig:LJ-CST} and~\ref{fig:LJ-CMS-CST-BvD} are related to this large NLO effect in the jet impact factor. Due to that, $C_0$ is
presumably exceedingly suppressed, thus leading to an average $\cos \varphi$ larger
than one for the lower rapidity values. In the larger rapidity region in
\textit{CASTOR-jet} configuration this issue is not present, since here the dominant part of the BFKL result originates from the iteration of BFKL kernel, and the role of NLO correction to the impact factors is less important.   
In order to qualitatively describe a processes with \textit{CASTOR-jet} kinematics, one
needs to develop a method to resum both BFKL energy logs and threshold double
logs. It is a very interesting problem, which goes however beyond the scope
of the present paper. Here we performed additional calculations for \textit{CASTOR}
kinematics with an enlarged rapidity interval for the detected $\Lambda$ hyperon and the jet, namely $|y_\Lambda| < 3$ and $-7.6 < y_J < -4.2$.
Our new plots in Figs.~\ref{fig:LJ-CST-lyb} and~\ref{fig:LJ-CST-lyb-BvD} with enlarged rapidity ranges show a considerable attenuation of the excess of $R_{10}$ in the low-$\Delta Y$ region. This occurs since in this case the effective value of the partonic $x$ on the jet side is shifted towards the region of smaller $x$, thus decreasing the role of threshold double logs.

\section{Closing statements}
\label{summary}

By proposing the inclusive detection of $\Lambda$ hyperons (with possible associated-jet emission) in the forward kinematic ranges of the LHC, we enriched the selection of semi-hard processes which can serve as a testing ground for the high-energy resummation, necessary ingredient in the investigation of the high-energy limit of strong interactions.
We performed a full NLA BFKL analysis of cross sections and azimuthal-angle correlations between the detected objects, finding in the dependence of these quantities on the final-state rapidity distance the usual onset of the high-energy dynamics. We proved that cross sections for {\LL} and {\LJ} production channels are from one to three orders of magnitude lower with respect to the ones typical of the light-charged di-hadron and hadron-jet reactions, respectively. This allows for an easier comparison with experimental data, by suppressing the contamination of the so-called minimum-bias events. In view of these results, we suggest experimental collaborations to consider the inclusion of the $\Lambda$-hyperon detection in the analyses of hadron(-jet) production in the LHC kinematic ranges sensitive to high-energy resummation physics.

\section*{Acknowledgments}

We thank Marco Radici for useful discussions on the physics of $\Lambda$ hyperons.
F.G.C. acknowledges support from the Italian Ministry of Education, Universities and Research under the FARE grant ``3DGLUE'' (n. R16XKPHL3N) and from the INFN/NINPHA project.
A.P. acknowledges support from the INFN/QFT@COLLIDERS project.

\end{document}